\documentclass[prd,aps,showpacs,preprintnumbers,amssymb]{revtex4}
\usepackage{epsf}

\begin{document}

\title{%
\hfill{\normalsize\vbox{%
\hbox{\rm }
 }}\\
{
Probing the substructure of  $f_0(1370)$
}}

\author{Amir H. Fariborz,
$^{\it \bf a}$~\footnote[1]{Email:
 fariboa@sunyit.edu}}

\author{Azizollah Azizi
$^{\it \bf b}$~\footnote[5]{Email:
   azizi@shirazu.ac.ir }}

\author{Abdorreza Asrar
$^{\it \bf b}$~\footnote[2]{Email:
   ar\_asrar@shirazu.ac.ir   }}

\affiliation{$^ {\bf \it a}$ Department of
Mathematics and Physics,
State University of New York, Polytechnic Institute, Utica, NY 13502, USA.}

\affiliation{$^ {\bf \it b}$ Department of Physics,  Shiraz University, Shiraz, Iran, }

\date{\today}

\begin{abstract}
Within an effective nonlinear chiral Lagrangian framework the substructure of $f_0(1370)$ is studied.
The investigation is conducted in the context of a  global picture of scalar mesons in which the importance of the  underlying connections  among scalar mesons below and above 1 GeV is recognized and implemented.    These connections are due to the mixings among various quark-antiquarks, four-quarks and glue components and play a central role in understanding the properties of scalar mesons.   Iterative Monte Carlo simulations are first performed on the 14-dimensional parameter space of the model and sets of points in this parameter space (the global sets) that give an overall agreement with all experimental data on mass spectrum, various decay widths and decay ratios of all isosinglet scalar states below 2 GeV are determined. Then within each global set, subsets that give closest agreement for the properties of $f_0(1370)$ are studied. Unlike the properties of other isosinglet states that show a range of variation within  each global set,  it is found that there is a clear signal for $f_0(1370)$  to be  predominantly a quark-antiquark state with a substantial  $s {\bar s}$ component, together with small  remnants of four-quark and glue components.
\end{abstract}


\maketitle

\section{Introduction}
The internal structure  of light scalar mesons has continued to challenge the quark model and QCD for several decades \cite{pdg}. The light and inverted mass spectrum for the lowest scalars does not obey a simple quark-antiquark picture which is known to work reasonably well for other spins such as pseudoscalars and vectors.   The MIT bag model of Jaffe \cite{Jaf},  which is based on a diquark-antidiquark picture,  explains why these states are lighter than expected and provides a natural template for their inverted mass spectrum.    While the states above 1 GeV are expected to be closer to quark-antiquark states,   nevertheless they too show deviations from such a simple scenario.     Particularly,  the case of isosinglet scalars are more involved than other isospin channels for they mix with glue in addition to the two- and four-quark components.     A wide range of investigations on scalar mesons (that include lattice QCD, QCD sum-rules, chiral models and effective theories) can be found in the literature \cite{Weinberg_13}-\cite{Achasov:1995mk} (some of which \cite{Janowski:2013uga}-\cite{Achasov:1995mk} have $f_0(1370)$ as one of their focus points).

Important information on scalars are obtained in various pseudoscalar scatterings such as in $\pi\pi$ channel for studies of the isosinglet states [particularly $f_0(500)$, or sigma ($\sigma$)];  in $\pi K$ channel for the studies of the isodoublet states $K_0^*(800)$ [or kappa ($\kappa$)] and $K_0(1430)$;  and in $\pi \eta$ channel for the studies of the isovector states $a_0(980)$ and $a_0(1450)$.   Chiral Lagrangians provide an effective framework for investigating pseudoscalar interactions.   Particularly,  chiral perturbation theory  \cite{ChPT} provides a systematic approach to studies of pion interactions near threshold.  In this approach,  pions are the main fields of interest and therefore the  heavier fields of vectors and scalars are integrated out.     However, for the purpose of exploring the properties of scalar mesons, which are outside the immediate focus of chiral perturbation theory,  it is natural to explicitly keep the scalar meson fields in the Lagrangian instead of integrating them out.     Two suitable frameworks,  that are the foundation of the present study,  are the linear sigma model \cite{3flavor,Ds_decays,LsM_mmp_eta3p,mixing_pipi,LsM_Maple,LsM_scatt_length,LsM_gauged,bfjpss09,global,07_FJS2,07_FJS4,07_FJS1,07_FJS3,05_FJS,05_FJS2,LsM,SU} as well as the nonlinear chiral Lagrangian models that include scalar fields \cite{SS,BFSS1,BFSS2,99FS,pieta,Blk_rad,e3p,06_F,Far_IJMPA,04_F,Mec}.     In such model buildings,  the main guiding principles are the well-known chiral symmetry and its breakdown,  isospin symmetry (and in relevant processes its breakdown), the U(1)$_{\rm A}$ axial anomaly,  and the main assumptions that  need to be made are related to modeling the QCD vacuum as well as the potential.        The choice between linear versus nonlinear is a matter of the processes to be investigated and the information to be extracted, nevertheless, they are overall complementary.   Prior works by one of the authors  within the linear and the nonlinear models (some of which listed in refs. \cite{3flavor}-\cite{Mec}) have indeed shown a consistent  pattern for the scalar mesons.   Specifically,  the properties of sigma and kappa  extracted in the nonlinear model in \cite{SS} and \cite{BFSS1} respectively,  are quantitatively close to those found within the linear sigma model \cite{LsM}; the four-quark nature of light scalar mesons below 1 GeV studied within the nonlinear model in \cite{BFSS2} are consistent with the results within the linear model in \cite{LsM};  and the underlying mixing patterns among the quark-antiquark and the four-quark components of scalars below and above 1 GeV studied in \cite{Mec} are consistent with similar patterns studied in detail within the generalized linear sigma model of refs. \cite{global,07_FJS2,07_FJS4,07_FJS1,07_FJS3,05_FJS}.

The approach we take in this paper is within the nonlinear model.
In the context of the nonlinear chiral Lagrangian of refs.
\cite{SS,BFSS1,BFSS2,99FS,pieta,Blk_rad,e3p},
several
low-energy processes that probe light scalar mesons are
investigated. In order to describe the experimental data
within this framework, there is a need for a
$\sigma$ and a $\kappa$ in the analysis of $\pi
\pi$ \cite{SS} and $\pi K$ \cite{BFSS1} scattering,
respectively.
Motivated by the evidence for a $\sigma$ and a $\kappa$,
and taking into account other experimentally
well-established scalars [the $f_0(980)$ and the
$a_0(980)$] a  possible
classification of these scalars (all below 1 GeV)  into a
nonet is studied in \cite{BFSS2} and it is  shown that there exists a unique choice of the free
parameters of this model which in addition to describing
the $\pi \pi$ and $\pi K$ scattering amplitudes, well
describes the experimental measurements for several decays,  such as, for example, the $\eta'
\to \eta \pi \pi $ decay \cite{99FS}.  The insight into the quark
substructure is obtained  through the mixing patterns between the properties of the isosinglets [$\sigma$ and $f_0(989)$]. The best value for
the isosinglet scalar mixing angle found in \cite{BFSS2} is clearly consistent with
an ideally mixed ${\bar q} {\bar q} q q$ assignment of the MIT bag model \cite{Jaf}).
Various  predictions within the framework of ref. \cite{BFSS2} are in close agreement with other experimental or theoretical works.  These include the 70 MeV estimate of
the total decay width of $a_0(980)$ in \cite{99FS} which is
confirmed  experimentally \cite{Teige}; the four-quark nature of light scalars probed in radiative $\phi$ decays
\cite{Ach,Blk_rad}; and the prediction  of sigma meson in agreement with experimental
analysis \cite{CLEO,07_D}.
Other low-energy processes in which scalar mesons are expected to play important roles  are studied within this framework including the
$\pi \eta$ scattering \cite{pieta}; and the isospin violating decays $\eta,\eta' \rightarrow 3 \pi$ \cite{e3p}.
Overall, the framework of \cite{BFSS2} has resulted in a coherent description for the physics of scalar mesons below 1 GeV.

The first step in extending the  framework of \cite{BFSS2} to include the scalars above 1 GeV (in addition to those below 1 GeV) was done in ref. \cite{Mec} in which two scalar meson nonets (a quark-antiquark nonet and a four-quark nonet) were introduced and the properties of $I=1/2$  and $I=1$  scalar mesons based on an underlying mixing among the two- and four-quark nonets were studied.     The idea that was introduced in \cite{Mec}  is rather simple, but quite effective:   Assuming that   there is  a four-qaurk scalar meson nonet below 1 GeV (which was proposed in MIT bag model \cite{Jaf} and has been since supported by many independent works), as well as a quark-antiquark scalar meson nonet above 1 GeV (which is expected to be a reasonable template for some of the scalars above 1 GeV),  then it is natural to investigate whether some of the observed deviations in mass and decay widths of the scalars above 1 GeV can be related to a mixing among these two- and four-quark nonets.    It was shown in \cite{Mec} how such a mixing leads to a ``level-repulsion'' that explains: (a)  why scalars below 1 GeV are so light, and   (b) why there are unexpected deviations in mass spectrum and decay widths of the $I=1/2,1$ scalars above 1 GeV (a brief review of this mixing mechanism is given in the Appendix).

Further steps in extending the framework of \cite{Mec} to include the $I=0$ states above 1 GeV were taken in \cite{Far_IJMPA,04_F,06_F} in which various additional mixings with glue are also present,  and as a result,  make the analysis considerably more challenging.   Specifically,  in \cite{Far_IJMPA,04_F} a preliminary  study of the mass spectrum as well as various  decay widths and decay ratios  of isosinglet scalars  were given, however, for simplicity, an uncorrelated analysis was performed
(the mass matrix for this system is a $5 \times 5$  matrix of eight a priori unknown parameters  and the scalar-pseudoscalar-pseudoscalar vertices depend on six coupling parameters,  therefore at first it may seem that the mass spectrum and the decay analyses are uncorrelated, but since the physical states that appear in the decay analysis are obtained by diagonalizing the mass matrix,  the decay analysis implicitly depends on the mass matrix parameters and consequently this establishes a correlation between these two sets of parameters).
In the work of \cite{06_F} the effect of  large experimental uncertainties on some of the scalar masses on determination of quark and glue components were studied in detail. In the present work,  we use an iterative Monte Carlo method (developed by the authors) to extend the works of \cite{Far_IJMPA,04_F} to a correlated analysis of the mass spectrum, decay widths and decay ratios of all $I=0$ scalars below 2 GeV  and  simultaneously examine the 14 unknown parameters in the mass and interaction parts of the Lagrangian.    This results in  estimating the substructure of all $I=0$ scalar states below 2 GeV, which are, in general,  rather sensitive to the experimental inputs and the physical conditions imposed.   However,   among the five isosinglet states [$f_0(500)$,  $f_0(980)$,  $f_0(1370)$,  $f_0(1500)$ and $f_0(1710)$],   the exception is $f_0(1370)$ which is found in our simulations to exhibit the most stable substructure and to be dominantly a  quark-antiquark (mainly $s {\bar s}$) state, further supporting prior findings \cite{Far_IJMPA,06_F}.

After a brief review of the model in Sec. II,  we set up the numerical work in Sec. III followed by results in Sec. IV and a summary in Sec. V.

\section{Brief review of the model}
The scalar properties are typically probed in various pseudoscalar scatterings or in their decay channels to pseudoscalars.    Therefore the light pseudoscalars ($\pi$, $K$ and $\eta/\eta'$) are essential ingredients in models  for investigating scalar mesons, and the model we are using in this work is no exception.  It will contain the psedudoscalars below 1 GeV as well as scalar mesons below and above 1 GeV.   In certain processes such as $\pi\pi$ scattering, the vector mesons also contribute  and in theose cases vectors are added too.    The leading pseudoscalar Lagrangian density is (see \cite{BFSS1})
\begin{equation}
{\cal L}_\phi= - {F_\pi^2\over 8} {\rm Tr} \left( \partial_\mu U \partial_\mu U^\dagger \right) + {\rm Tr}\left[ {\cal B} \left( U + U^\dagger \right)\right],
\label{Lphi}
\end{equation}
with
\begin{equation}
U = e^{2 i \phi/ F_\pi},
\end{equation}
where $\phi$ is the conventional matrix of pseudoscalar fields and $F_\pi =0.131$ GeV is the pion decay constant. The second term is the  symmetry breaking term with ${\cal B} = \mathrm{diag} (B_1, B_1, B_3)$, with $B_1 = m_\pi^2 F_\pi^2 /8$ and $B_3=F_\pi^2(m_K^2 - m_\pi^2/2)/4$.    Moreover,  the U(1)$_{\rm A}$ breaking terms induced by instanton effects need to be added to the Lagrangian density (\ref{Lphi}) to generate the $\eta'$ mass
\begin{equation}
{\cal L}_{\eta'} = { {{\tilde c}\over 576}} \left[ {\rm ln} \left( { {{\rm det} U}\over {{\rm det} U^\dagger} } \right)\right]^2 +\cdots ,
\label{U1A}
\end{equation}
where ${\tilde c}$ is a constant proportional to the $\eta'$ mass (the dots represent additional terms given in Eq. (2.12) of \cite{SSW_1993}).   Note that the  functional form of Eq. (\ref{U1A}), expressed in terms of $\ln$ and $\det$ functions,  schematically shows that chiral SU(3)$_{\rm L} \times$  SU(3)$_{\rm R}$ symmetry is maintained while the U(1)$_{\rm A}$ is broken to generate a mass term for $\eta'$.

Under chiral transformation
\begin{equation}
U\rightarrow U_L U U_R^\dagger.
\label{U_trans}
\end{equation}
The nonlinear model is derived by integrating out the heavy fields of scalar mesons and gives a  convenient framework for investigating the Goldstone bosons interactions near the threshold.     However,  for our present objective of exploring the properties of scalar mesons we need to reintroduce the scalars back into the Lagrangian.    One way of course is to start from the linear sigma model, form chiral nonets and study the interactions.   This is done in refs. \cite{3flavor,Ds_decays,LsM_mmp_eta3p,mixing_pipi,LsM_Maple,LsM_scatt_length,LsM_gauged,bfjpss09,global,07_FJS2,07_FJS4,07_FJS1,07_FJS3,05_FJS,05_FJS2,LsM,SU} and involves the heavy pseudoscalars $\pi(1300)$, $K(1460)$ and several etas [$\eta(1290)$, $\eta(1405)$ and $\eta(1760)$] which are related to the chiral partners of heavy scalar mesons.    An advantage of the nonlinear approach is that it simplifies the framework by not bringing in the heavy pseudoscalars which are actually not needed in the mass spectrum and decay analyses of the present work.      The disadvantage of the nonlinear model is that it looses contact with the ``nuts and bolts'' of chiral symmetry and its spontaneous breakdown via QCD vacuum condensates, which is so transparently traced in the linear model.   Nevertheless, for the purpose of analyzing the mass spectrum and decay channels of isosinglet states, the nonlinear model is reasonably  effective.   As stated before, in general,  various analyses within the nonlinear model in \cite{e3p,Blk_rad,pieta,99FS,BFSS2,BFSS1,SS,06_F,Far_IJMPA,04_F,Mec} have been consistent  and complementary to those in the linear model \cite{3flavor,Ds_decays,LsM_mmp_eta3p,mixing_pipi,LsM_Maple,LsM_scatt_length,LsM_gauged,bfjpss09,global,07_FJS2,07_FJS4,07_FJS1,07_FJS3,05_FJS,05_FJS2,LsM,SU}.

It is noted that under chiral transformation the field $\xi$ defined by $U=\xi^2$ transforms as
\begin{equation}
\xi\rightarrow U_L \xi {\cal K}^\dagger \left(\phi,U_L,U_R\right) = {\cal K}\left(\phi,U_L,U_R\right)\xi U_R^\dagger,
\end{equation}
where ${\cal K}$ is defined in the above equation.   Then it is easy to show that under chiral transformation the object
\begin{equation}
p_\mu = {i\over 2}\left( \xi \partial \xi^\dagger - \xi^\dagger\partial \xi\right),
\end{equation}
transforms as
\begin{equation}
p_\mu \rightarrow {\cal K} p_\mu {\cal K}^\dagger.
\label{p_trans}
\end{equation}
To reintroduce the scalars back into the nonlinear Lagrangian, it was considered in \cite{BFSS2}  that the scalar nonets were made of ``constituent'' quarks and transform in the same way as (\ref{p_trans})
\begin{equation}
N \rightarrow {\cal K} N {\cal K}^\dagger,
\label{N_trans}
\end{equation}
where $N$ is a four-quark scalar meson nonet \cite{BFSS2}
which is defined in terms of diquark-antidiquark fields
\begin{eqnarray}
Q_a &=& \epsilon_{a b c} {\bar q}^b {\bar q}^c, \nonumber \\
{\bar Q}^a &=& \epsilon^{a b c} q_b q_c,
\end{eqnarray}
according to
\begin{equation}
N_a^b
\propto Q_a  {\bar Q}^b =
                         \left[
                         \begin{array}{c c c}
{\bar d} {\bar s} d s & {\bar d} {\bar s} s u & {\bar d} {\bar s} u d \\
{\bar s} {\bar u} d s & {\bar s} {\bar u} s u & {\bar s} {\bar u} u d \\
{\bar u} {\bar d} d s & {\bar u} {\bar d} s u & {\bar u} {\bar d} u d \\
                           \end{array}
                         \right].
\end{equation}
The transformations (\ref{p_trans}) and (\ref{N_trans}) allow writing the Lagrangian terms for scalar fields as well as the scalar-pseudoscalar-pseudoscalar interaction terms which involve nonlinear pion terms (for details see Appendix B of \cite{BFSS2}).     Similarly, a conventional quark-antiquark nonet ${N'}_a^b \propto q_a {\bar q}^b$ (that has exactly the same transformation property, and hence can mix with $N$) can be added
\begin{equation}
N' \rightarrow {\cal K} N' {\cal K}^\dagger.
\label{Np_trans}
\end{equation}
This was introduced in ref. \cite{Mec} in which the case of $I=1/2,1$ states were studied in detail.

Here,  the relevant terms for our analysis are the scalar mass terms and the scalar-pseudoscalar-pseudoscalar interaction terms which are extracted from the general chiral invariant Lagrangian  of ref. \cite{BFSS2} together with the mixing mechanism  of \cite{Mec}.  In addition a scalar field $G$ that represents  the effective field of a scalar glueball and is relevant to the study of isosinglet states is also added \cite{Far_IJMPA}.    Identification of $G$ with scalar glueball,  which is discussed in detail in \cite{FAA_1500},  is based on the results of various fits to experiment which show that, for example: (i) $f_0(1500)$ and $f_0(1710)$ are the only two states that contain a high content of this field; (ii) $G$ couples strongly to pseudoscalar-pseudoscalar channels that involve $\eta'$; and  (iii) the mass of $G$ is determined to be in the range of 1.5-1.8 GeV consistent with lattice QCD estimates.  These observations are all consistent with identifying field $G$ with an scalar glueball.   The mass and interaction parts of the Lagrangian are
\begin{eqnarray}
{\cal L}_{\rm mass}^{I=0}  &=& {\cal L}_{\rm mass}^{I=1/2,1}
- c {\rm Tr}(N){\rm Tr}(N)
- d {\rm Tr}(N) {\rm Tr}(N{\cal M})
\nonumber \\
&&- c' {\rm Tr}(N'){\rm Tr}(N')
- d' {\rm Tr}(N') {\rm Tr}(N'{\cal M})
- g G^2
- \rho {\rm Tr} (N) {\rm Tr} (N')
- e G {\rm Tr} \left( N \right)
- f G {\rm Tr} \left( N' \right),\label{L_mass_I0} \\
{\cal L}_{\rm int}^{I=0} &=& {\cal L}_{\rm int}^{I=1/2,1} + B {\rm Tr} \left( N \right) {\rm Tr} \left({\partial_\mu}\phi
{\partial_\mu}\phi \right)
+ D {\rm Tr} \left( N \right) {\rm Tr}
\left({\partial_\mu}\phi \right)  {\rm Tr} \left( {\partial_\mu}\phi
\right)
+ B' {\rm Tr} \left( N' \right) {\rm Tr} \left({\partial_\mu}\phi
{\partial_\mu}\phi \right)
\nonumber\\
&& + D' {\rm Tr} \left( N' \right) {\rm Tr}
\left({\partial_\mu}\phi \right)  {\rm Tr} \left( {\partial_\mu}\phi
\right)
+ E G  {\rm Tr} \left({\partial_\mu}\phi
{\partial_\mu}\phi \right)
+ F G  {\rm Tr}
\left({\partial_\mu}\phi \right)  {\rm Tr} \left( {\partial_\mu}\phi
\right),
\label{L_int_I0}
\end{eqnarray}
where the matrix ${\cal M} = {\rm diag} (1,1,x)$ with $x$ being the ratio of
the strange to non-strange quark masses, and ${\cal L}_{\rm mass}^{I=1/2,1}$ and ${\cal L}_{\rm int}^{I=1/2,1}$ are the mass terms and the scalar-pseudoscalar-pseudoscalar interaction terms relevant to the $I=1/2,1$ states studied in ref. \cite{Mec} that also contribute to the $I=0$ states.  Therefore, part of the Lagrangian of isosinglet states is constrained by the properties of $I=1/2,1$ scalar mesons.
All free parameters in $I=1/2,1$ parts are determined in fits to the mass spectrum and decay properties of $I=1/2,1$ states in \cite{Mec} (these parts are briefly presented in Appendix A).
It is also shown in the Appendix A that the mixing term between the two- and the four-quark nonets $N$ and $N'$ is similar to the instanton contribution to the scalar sector which is studied in the literature \cite{Klempt_1995,Minkowski_99,Dmitrasinovic_96} and is suggested to be important for the isosinglet scalar states. We see  that the investigation of isosinglet states is not independent of the isodoublets and isovectors and indeed is constrained by them.
The rest of the Lagrangian density only contribute to the $I=0$ states and is considerably more complex than the $I=1/2,1$ parts due to the fact that there are internal mixings of two isospin zero combinations within each nonet $N$ and $N'$, as well as the mixings of these combinations with a scalar glueball, and as a result, there are more parameters to keep track of and the  mixing matarix is $5\times 5$ as opposed to $2\times 2$ mixings in the case of isosinglets and isovectors. Initial studies of the $I=0$ sector are given in \cite{Far_IJMPA,04_F,06_F}  and will be generalized in the present investigation.  There are altogether 14 unknown parameters, eight of these ($c$, $d$, $c'$, $d'$, $g$, $\rho$, $e$ and $f$) only contribute to the isosinglet $5\times 5$ mass matirx (in addition,  there are also contributions to the $I=0$ mass matrix coming from ${\cal L}_{\rm mass}^{1/2,1}$),   and the remaining six parameters ($B$, $D$, $B'$, $D'$, $E$ and $F$) contribute to the scalar-pseudoscalar-pseudoscalar coupling constants (there are also contributions to the interaction vertices coming from ${\cal L}_{\rm int}^{1/2,1}$).

Our convention for the mass matrix is as follows:  Altogether, there are five $I=0$ combinations,  two in the four-quark nonet [$N_3^3$ and $(N_1^1 + N_2^2)/\sqrt{2}$], two in the quark-antiquark nonet [${N'}_3^3$ and $({N'}_1^1 + {N'}_2^2)/\sqrt{2}$] and a scalar glueball ($G$). We organize these components into a column matrix
\begin{equation}
{\bf F}_0 =
       \begin{array}{l}
                         \left(
                         \begin{array}{c}
                              N_3^3 \\
                             (N_1^1 + N_2^2)/\sqrt{2}\\
                              {N'}_3^3 \\
                             ({N'}_1^1 + {N'}_2^2)/\sqrt{2}\\
                              G
                         \end{array}
                         \right)
       \end{array}
                         =
                         \left(
                         \begin{array}{c}
      						f_0^{NS}\\
                            f_0^S\\
                            f_0^{\prime S}\\
                            f_0^{\prime NS} \\
                              G
                         \end{array}
                         \right)
\propto
       \begin{array}{l}
                         \left(
                         \begin{array}{c}
      {\bar u}{\bar d} u d\\
     ({\bar d}{\bar s} d s + {\bar s}{\bar u} s u) /\sqrt{2}\\
                              s {\bar s}\\
                             (u {\bar u} + d {\bar d})/\sqrt{2}\\
                              \alpha_s G^{\mu\nu} G_{\mu\nu}
                         \end{array}
                         \right),
       \end{array}
\label{SNS_def}
\end{equation}
where the corresponding quark substructures are shown on the right and $S$ and $NS$ stand for strange and non-strange. Then the mass terms are organized into the mass matrix
\begin{equation}
 - {1\over 2} {\tilde {\bf
F}}_0 {\bf M}^2 {\bf
F}_0 =
- {1\over 2} {\tilde {\bf F}} {\bf M}_\mathrm{diag}^2 {\bf F},
\label{L_mass_mix}
\end{equation}
where ${\bf F}$ contains the five isosinglet physical fields
\begin{equation}
{\bf F}  =
       \begin{array}{c}
                        \left( \begin{array}{c}
                                f_0(500)\\
                                f_0(980)\\
                                f_0(1370)\\
                                f_0(1500)\\
                                f_0(1710)
                                \end{array}
                         \right)
= K_F^{-1} {\bf F}_0,
       \end{array}
\label{K_def}
\end{equation}
and $K_F^{-1}$ is the rotation matrix that converts the quark and glue basis into the physical basis.

The unknown parameters $c$ and $d$ induce ``internal'' mixing
between the two $I=0$ flavor combinations [$(N_1^1 + N_2^2)/\sqrt{2}$ and
$N_3^3$] of nonet $N$. Similarly,  $c'$ and $d'$ play the same
role in nonet $N'$.
Parameters $c,d,c'$ and $d'$ do not contribute to
the mass spectrum of the $I=1/2$ and $I=1$ states.     The ``external'' mixing between nonets $N$ and $N'$ (the $\rho$ term), the glueball mass term (the $g$ term), and the glueball  mixing terms with nonets $N$ and $N'$ (the $e$ and $f$ terms) are also given in Eq.~(\ref{L_mass_I0}).
Parameters $B$ and $D$ are unknown coupling constants describing the
coupling of the four-quark nonet $N$ to the pseudoscalars, parameters
$B'$ and $D'$  are couplings of $N'$ to the pseudoscalars, and parameters $E$ and $F$
describe the coupling of a scalar glueball to the pseudoscalar mesons.

After diagonalization of the mass matrix and rotation of the quark and glue basis to the physical basis, the interaction Lagrangian (\ref{L_int_I0}) can be rewritten as:
\begin{eqnarray}
- {\cal L}_\mathrm{int} &=&
{1\over \sqrt{2}} \gamma_{\pi\pi}^i \, {\bf F}_i \partial_\mu \pi \cdot
\partial_\mu
\pi
+ {1\over \sqrt{2}} \gamma_{KK}^i \, {\bf F}_i \partial_\mu {\bar K}
\partial_\mu K
\nonumber \\
&& + \gamma_{\eta\eta}^i \, {\bf F}_i \partial_\mu \eta
\partial_\mu \eta
+ \gamma_{\eta\eta'}^i \, {\bf F}_i \partial_\mu \eta
\partial_\mu \eta'
+ \gamma_{\eta'\eta'}^i \, {\bf F}_i \partial_\mu \eta'
\partial_\mu \eta',
\label{L_int_I0_b}
\end{eqnarray}
where $\gamma^i_{ss'}$ is the coupling of the $i$-th isosinglet scalar [with $i=1\ldots 5$ corresponding to the physical states $f_0(500) \ldots f_0(1710)$; see Eq.~(\ref{K_def})] to
pseudoscalars $s$ and $s'$, and is given by
\begin{equation}
\gamma^i_{ss'} = \sum_j \left( \gamma_{ss'} K\right)_{ji},
\end{equation}
with  $K$ defined in  ({\ref{K_def}) and
$
\gamma_{ss'} =
{\rm diag}
\left(\gamma^{NS}_{ss'}, \gamma^S_{ss'}, \gamma'^S_{ss'},\gamma'^{NS}_{ss'},\gamma^G_{ss'}
\right),
$
in which the diagonal elements are the couplings of the
pseudoscalars $s$ and $s'$ to the quark and glue basis $f_0^{NS}, f_0^S, {f'}_0^S, {f'}_0^{NS}$, $G$
[defined in (\ref{SNS_def})], respectively.
The diagonal elements for all decay channels $ss'$ are given in \cite{Far_IJMPA}.
In next two sections we give the details of our numerical determination of the 14 free parameters in the Lagrangian density.

\section{Setting up the numerical analysis}
There are 14 unknown parameters in the $I=0$ part of the Lagrangian density that we need to determine by incorporating appropriate experimental data on the mass spectrum as well as the appropriate decay widths and decay ratios.      These can be divided into a six-dimensional parameter space ($B$, $D$, $B'$, $D'$, $E$ and $F$) that only affect the scalar-pseudoscalar-pseudoscalar coupling constants, and an eight-dimensional parameter space ($c$, $d$, $c'$, $d'$, $g$, $\rho$, $e$ and $f$) that both directly enter into the $5\times5$ mass matrix, as well as indirectly enter in the calculation of decay widths and decay ratios through the rotation matrix $K$ that rotates the ``bare'' bases into the physical bases.    As a result,  determining these two groups of parameters independent of each other is only an approximation and the exact determination requires a simultaneous 14-parameter fit.    In works of refs. \cite{Far_IJMPA,04_F,06_F}, as a preliminary approach,   these two parameter spaces were studied in separate fits in some details.  Here we generalize those separate fits into one simultaneous fit using an iterative Monte Carlo algorithm.

Since the experimental status of scalar mesons is not yet firmly established, different  existing data do  not always overlap.     To reduce uncertainties that stem from unestablished experimental data, we incorporate several sources of input.    We have collected the inputs that we use into two Tables \ref{quantities1} and \ref{quantities2}.
The experimental inputs in Table \ref{quantities1} are divided into three groups:   the masses and several decay ratios and  decay widths.     Altogether,  23 inputs are displayed in Table \ref{quantities1}.   In principle one might think about selecting a subset  14 out of these 23 experimental data and examine whether a determined system of 14 equations in 14 unknowns might be formed.    However, we do not seek to solve a mathematical system, because even if such a determined system exists, solving such a highly nonlinear system and finding all distinct solutions, can be at the expense of pushing the model predictions away from other experimental quantities not included in the chosen set of 14 inputs.  Our objective is to explore the underlying mixings among various two-quark, four-quark and glue components, in order to achieve a global understanding of all $I=0$ states.
This objective is sometimes at the expense of individual accuracies,  at least at the present approximation of the model.      Therefore,  we aim to determine the 14 Lagrangian parameters such that we get a reasonable agreement with the experimental data displayed in Table \ref{quantities1}.  We use the inputs of Table \ref{quantities1} in two ways:  In our global fit I, we exclude the two partial decay widths of $f_0(1370)$ and input the rest of 21 quantities.  This is because of the fact that even though these two decay widths are listed in PDG \cite{pdg} they are not used in any averaging and therefore we would like to test their importance.  In global fit II we include all the 23 quantities (including the two decay widths ignored in the global fit I). Our global fit III is obtained from the inputs of Table \ref{quantities2} which does not include the decay ratios provided by the WA102 collaboration \cite{WA102} displayed in Table \ref{quantities1}.

\begin{table}[t]
	\begin{center}
		\begin{tabular}{|c|c|c|}
			\hline
			Short notation & Quantity &  Experimental value [ref.] \\ \hline
			$m_1$  & $m[f_0(500)]$& $400-550$  {\rm MeV} \cite{pdg} \\ \hline
			$m_2$ & $m[f_0(980)]$ & $990 \pm 20$ {\rm MeV} \cite{pdg}\\ \hline
			$m_3$ & $m[f_0(1370)]$ & $1312$  {\rm MeV} \cite{WA102} \\ \hline
			$m_4$ & $m[f_0(1500)]$&$ 1502$  {\rm MeV} \cite{WA102} \\ \hline
			$m_5$ & $m[f_0(1710)]$& $1727$  {\rm MeV} \cite{WA102} \\ \hline \hline
			$\Gamma^3_{\frac{\pi\pi}{KK}}$&$\frac{\Gamma[f_0(1370)\rightarrow\pi\pi]}{\Gamma[f_0(1370)\rightarrow K\bar{K} ]}$&$2.17  \pm  0.9$ \cite{WA102}\\ \hline
			 $\Gamma^3_{\frac{\eta\eta}{KK}}$&$\frac{\Gamma[f_0(1370)\rightarrow\eta\eta]}{\Gamma[f_0(1370)\rightarrow K\bar{K} ]}$&$0.35  \pm  0.30$  \cite{WA102}\\ \hline
			 $\Gamma^4_{\frac{\pi\pi}{\eta\eta}}$&$\frac{\Gamma[f_0(1500)\rightarrow\pi\pi]}{\Gamma[f_0(1500)\rightarrow \eta\eta ]}$& $5.56  \pm 0.93$ \cite{WA102}\\ \hline
			$\Gamma^4_{\frac{KK}{\pi\pi}}$&$\frac{\Gamma[f_0(1500)\rightarrow K\bar{K}]}{\Gamma[f_0(1500)\rightarrow \pi\pi ]}$& $0.33  \pm  0.07$  \cite{WA102}\\ \hline
			$\Gamma^4_{\frac{\eta \eta'}{\eta \eta}}$&$\frac{\Gamma[f_0(1500)\rightarrow\eta \eta' ]}{\Gamma[f_0(1500)\rightarrow \eta\eta ]}$&  $0.53  \pm  0.23$ \cite{WA102}\\ \hline
			$\Gamma^5_{\frac{\pi\pi}{KK}}$&$\frac{\Gamma[f_0(1710)\rightarrow\pi\pi]}{\Gamma[f_0(1710)\rightarrow K\bar{K} ]}$& $0.20  \pm  0.03$  \cite{WA102}\\ \hline
			$\Gamma^5_{\frac{\eta \eta}{KK}}$&$\frac{\Gamma[f_0(1710)\rightarrow\eta\eta]}{\Gamma[f_0(1710)\rightarrow K\bar{K} ]}$&$ 0.48  \pm   0.19$ \cite{WA102}\\ \hline \hline
			$\Gamma^1_{\pi\pi}$&$\Gamma[f_0(500)\rightarrow\pi\pi]$&$ 400-700$ {\rm MeV} \cite{pdg}  \\ \hline
			$\Gamma^2_{\pi\pi}$&$\Gamma[f_0(980)\rightarrow\pi\pi]$& $ 40-100$ {\rm MeV} \cite{pdg}\\ \hline
			$\Gamma^3_{\pi\pi}$&$\Gamma[f_0(1370)\rightarrow\pi\pi]$&$(0.26 \pm 0.09)\times(230 \pm 15)$ {\rm MeV} \cite{Bugg:2007ja}\\ \hline
			$\Gamma^3_{KK}$&$\Gamma[f_0(1370)\rightarrow K\bar{K}]$&$(0.35 \pm 0.13)\times(230 \pm 15)$ {\rm MeV} \cite{Bugg:2007ja}\\ \hline
			$\Gamma^4_{\pi\pi}$&$\Gamma[f_0(1500)\rightarrow\pi\pi]$& $(0.349 \pm 0.023)\times(109 \pm 7)$ {\rm MeV} \cite{pdg}\\ \hline
			$\Gamma^4_{KK}$&$\Gamma[f_0(1500)\rightarrow K\bar{K}]$& $(0.086 \pm 0.010 )\times(109 \pm 7)$ {\rm MeV} \cite{pdg} \\ \hline
			$\Gamma^4_{\eta\eta}$&$\Gamma[f_0(1500)\rightarrow\eta\eta]$& $(0.051 \pm 0.009)\times(109 \pm 7)$ {\rm MeV}  \cite{pdg}\\ \hline
			$\Gamma^4_{\eta\eta\prime}$&$\Gamma[f_0(1500)\rightarrow\eta\eta\prime]$& $(0.019 \pm 0.008)\times(109 \pm 7)${\rm MeV}\cite{pdg}\\ \hline
			$\Gamma^5_{\pi\pi}$&$\Gamma[f_0(1710)\rightarrow\pi\pi]$& $(0.12 \pm 0.11)\times(220 \pm 40)$ {\rm MeV} \cite{oller}\\ \hline
			$\Gamma^5_{KK}$&$\Gamma[f_0(1710)\rightarrow K\bar{K}]$& $(0.36\pm 0.12)\times(220 \pm 40)$ {\rm MeV} \cite{oller}\\ \hline
			$\Gamma^5_{\eta\eta}$&$\Gamma[f_0(1710)\rightarrow \eta\eta]$& $(0.22 \pm 0.12)\times(220 \pm 40)$ {\rm MeV} \cite{oller} \\ \hline
		\end{tabular}
	\end{center}
\caption{Target quantities used to explore the 14 parameters of the Lagrangian in global fits I and II.  In global  fit I (II) the decay channels of $f_0(1370)$  are excluded (included).   The short notation for the quantities are defined in column one.}
\label{quantities1}
\end{table}

\begin{table}[t]
	\begin{center}
		\begin{tabular}{|c|c|c|}
			\hline
			Short notation & Quantity &  Experimental value [ref.] \\ \hline
			$m_1$  & $m[f_0(500)]$& $400-550$  {\rm MeV} \cite{pdg} \\ \hline
			$m_2$ & $m[f_0(980)]$ & $990 \pm 20$ {\rm MeV} \cite{pdg}\\ \hline
			$m_3$ & $m[f_0(1370)]$ & $1300 \pm 15$  {\rm MeV} \cite{Bugg:2007ja} \\ \hline
			$m_4$ & $m[f_0(1500)]$&$ 1505 \pm 6$  {\rm MeV} \cite{pdg} \\ \hline
			$m_5$ & $m[f_0(1710)]$& $1690 \pm 20$  {\rm MeV} \cite{oller} \\ \hline \hline
			$\Gamma^1_{\pi\pi}$&$\Gamma[f_0(500)\rightarrow\pi\pi]$&$ 400-700$ {\rm MeV} \cite{pdg}  \\ \hline
			$\Gamma^2_{\pi\pi}$&$\Gamma[f_0(980)\rightarrow\pi\pi]$& $ 40-100$ {\rm MeV} \cite{pdg}\\ \hline
			$\Gamma^3_{\pi\pi}$&$\Gamma[f_0(1370)\rightarrow\pi\pi]$&$(0.26 \pm 0.09)\times(230 \pm 15)$ {\rm MeV} \cite{Bugg:2007ja}\\ \hline
			$\Gamma^3_{KK}$&$\Gamma[f_0(1370)\rightarrow K\bar{K}]$&$(0.35 \pm 0.13)\times(230 \pm 15)$ {\rm MeV} \cite{Bugg:2007ja}\\ \hline
			$\Gamma^4_{\pi\pi}$&$\Gamma[f_0(1500)\rightarrow\pi\pi]$& $(0.349 \pm 0.023)\times(109 \pm 7)$ {\rm MeV} \cite{pdg}\\ \hline
			$\Gamma^4_{KK}$&$\Gamma[f_0(1500)\rightarrow K\bar{K}]$& $(0.086 \pm 0.010 )\times(109 \pm 7)$ {\rm MeV} \cite{pdg} \\ \hline
			$\Gamma^4_{\eta\eta}$&$\Gamma[f_0(1500)\rightarrow\eta\eta]$& $(0.051 \pm 0.009)\times(109 \pm 7)$ {\rm MeV}  \cite{pdg}\\ \hline
			$\Gamma^4_{\eta\eta\prime}$&$\Gamma[f_0(1500)\rightarrow\eta\eta\prime]$& $(0.019 \pm 0.008)\times(109 \pm 7)${\rm MeV}\cite{pdg}\\ \hline
			$\Gamma^5_{\pi\pi}$&$\Gamma[f_0(1710)\rightarrow\pi\pi]$& $(0.12 \pm 0.11)\times(220 \pm 40)$ {\rm MeV} \cite{oller}\\ \hline
			$\Gamma^5_{KK}$&$\Gamma[f_0(1710)\rightarrow K\bar{K}]$& $(0.36\pm 0.12)\times(220 \pm 40)$ {\rm MeV} \cite{oller}\\ \hline
			$\Gamma^5_{\eta\eta}$&$\Gamma[f_0(1710)\rightarrow \eta\eta]$& $(0.22 \pm 0.12)\times(220 \pm 40)$ {\rm MeV} \cite{oller} \\ \hline
		\end{tabular}
	\end{center}
	\caption{Target quantities used in global fit III.  \label{quantities2}}
\end{table}
We measure the goodness of the fits by the smallness of the parameter $\chi$ defined by
\begin{equation}
\chi\left( p_1\ldots p_{14} \right) = \sum_{i=1}^{N_q^{\rm exp}}
              \left|
              {
                { {\hat q}_i^{\rm exp} - q_i^{\rm theo}\left(p_1\ldots p_{14}\right)}
                               \over
                          {{\hat q}_i^{\rm exp}}
              }
               \right|,
\end{equation}
where $q_i^{\rm exp} = {\hat q}_i^{\rm exp} \pm \Delta q_i^{\rm exp}$ with $i=1\ldots N_q^{\rm exp}$ (for quantities that an experimental range is reported, $q_i^{\rm exp} =
q_{i, {\rm min}}^{\rm exp}\ldots q_{i, {\rm max}}^{\rm exp}$,  we take ${\hat q}_i^{\rm exp}$ to be the central value).  Our target quantities are  ${\hat q}_i^{\rm exp}$  which are also theoretically calculated by the model $q_i^{\rm theo}$ as a function of the 14 model parameters ($p_1\ldots p_8$ = $c$, $d$, $c'$, $d'$, $g$, $\rho$, $e$, $f$ and $p_9\ldots p_{14}$ = $B$, $D$, $B'$, $D'$, $E$ and $F$).
This results in a ``fixed target method'' in which the computation revolves around reproducing the central values of the experimental data, which here in this work is just one set.   This is to be contrasted with a ``moving target method'' in which every point within the experimental range is treated as a viable target and the computation spans over all possible target sets to find the best agreement with the model computation.
The guiding function $\chi$,  that was introduced in \cite{06_F},  has two important advantages that are suitable for our study of the underlying mixings and the global picture of scalar mesons:  (a) it gives each individual data an equal weight (unlike the conventional $\chi^2$ method); and (b) it can be used when dealing with different quantities, such as here that we input different types of experimental quantities of masses, decay widths and decay ratios.   To measure the goodness of our fits, we compare a  $\chi$ with its corresponding experimental value defined by
\begin{equation}
\chi^{\rm exp} =
\sum_{i=1}^{N_q^{\rm exp}}
              \left|
              {
                {\Delta q_i^{\rm exp}}
                               \over
                          {{\hat q}_i^{\rm exp}}
              }
               \right|.
\end{equation}
Since in most situations there is nothing unique about the experimental central values, we do not limit our simulations to just searches  for the best fit (or lowest $\chi$ value).  Instead, we adopt an inclusive process that accounts for the experimental uncertainties (expressed by $\chi^{\rm exp}$) by filtering out simulations that do not satisfy $\chi \le \chi^{\rm exp}$ condition.    This method in general leads to finding an acceptable set instead of just the best point which obviously is also included in the set (the method results in ``dispersive fits'' that reflect the experimental uncertainties, but for simplicity we refer to them as ``fits'' throughout this work).  In the moving target method,  for each choice of an experimental set, there is a corresponding $\chi^{\rm exp}$, and therefore, overall this method leads to a range for $\chi^{\rm exp}$.

For the details of our numerical analysis we will use several strategies and define the guiding function $\chi$ accordingly.  In general our guiding function contains three parts
\begin{equation}
\chi \left(p_1\ldots p_{14}\right) = \chi_m \left(p_1\ldots p_8\right) + \chi_\Gamma \left(p_1\ldots p_{14}\right) +
\chi_{(\Gamma/\Gamma)} \left(p_1\ldots p_{14}\right),
\label{chi_generic}
\end{equation}
where the three terms on the right refer to $\chi$ for mass, decay width and decay ratio, defined by
\begin{eqnarray}
\chi_m \left(p_1\ldots p_8\right) &=& \sum_{i=1}^5
              \left|
              {
                { {\hat m}_i^{\rm exp} - m_i^{\rm theo}\left(p_1\ldots p_8\right)}
                               \over
                          {{\hat m}_i^{\rm exp}}
              }
               \right|,
\nonumber \\
\chi_\Gamma\left(p_1\ldots p_{14}\right) &=& \sum_{i=1}^5
              \sum_\alpha
              \left|
              {
                { \left({\hat \Gamma}_\alpha^i\right)^{\rm exp} -
                   \left(\Gamma_\alpha^i\right)^{\rm theo}\left(p_1\ldots p_{14}\right)}
                               \over
                          {\left({\hat \Gamma}_\alpha^i\right)^{\rm exp}}
              }
               \right|,
\nonumber \\
\chi_{(\Gamma/\Gamma)}\left(p_1\ldots p_{14}\right) &=& \sum_{i=1}^5
\sum_\alpha \sum_\beta
              \left|
              {
                { \left({\hat \Gamma^i_{\alpha/\beta}}\right)^{\rm exp} -
                \left(\Gamma^i_{\alpha/\beta}\right)^{\rm theo}\left(p_1\ldots p_{14}\right)}
                               \over
                          { {\left( {\hat \Gamma}^i_{\alpha/\beta}\right)}^{\rm exp}}
              }
               \right|,
\label{m_DW_DR_def}
\end{eqnarray}
with short notations
\begin{eqnarray}
\Gamma_\alpha^i &=& \Gamma \left[f_i \rightarrow \alpha\right], \nonumber \\
\Gamma^i_{\alpha/\beta} &=& { {\Gamma\left[f_i \rightarrow \alpha\right]}\over {\Gamma
\left[f_i \rightarrow \beta\right]}},
\end{eqnarray}
where $i=1\ldots 5$ correspond to the five isosinglet scalars in ascending order of masses,  and $\alpha$ and $\beta$ are the two-body decay channels and in this work take values  $1\ldots 4$ which respectively correspond to the decay channels $\pi\pi$, $\pi K$, $\eta \eta$ and $\eta \eta'$ (note: it is understood that the summations run over relevant values of $\alpha$ and $\beta$ that are listed in Tables \ref{quantities1} and \ref{quantities2}).

\section{Results}
The investigation of the substructure of $f_0(1370)$ in the present work,  is based on first establishing  global relationships among all relevant scalars below and above 1 GeV and then zooming in on $f_0(1370)$.
With that in mind,  we first perform global fits of the model predictions to experimental data for the quantities given in Tables \ref{quantities1} and \ref{quantities2} and search within the 14-dimensional parameter space for sets of acceptable points that satisfy an overall agreement between model predictions and experiment.    Then within these acceptable sets, we further zoom in on properties of $f_0(1370)$.     This leads to our global fits and their further refinements in this section.

\subsection{Global fits}
To obtain the acceptable points within our 14d parameter space that provide an overall acceptable description of all relevant scalars,  we perform three global fits.   In global fit I, we exclude the two decay widths of $f_0(1370)$ from the 23 target list of Table \ref{quantities1} and in global fit II we include all the 23 inputs of that Table. Global fit III is obtained with the target values of Table \ref{quantities2}.

\subsubsection{Global fit I}

In global fit I,  where the partial decay widths of $f_0(1370)$ are excluded from the target inputs,  the guiding function for $\chi_\mathrm{I}$ is computed from (\ref{chi_generic}) in which $\chi_m$, $\chi_\Gamma$ and $\chi_{(\Gamma/\Gamma)}$ are obtained from (\ref{m_DW_DR_def}),  with the condition that in $\chi_\Gamma$ the decay widths of $f_0(1370)$ have been excluded (i.e. $i\ne 3$).   The $\chi_m$ and $\chi_{(\Gamma/\Gamma)}$ are those given in (\ref{m_DW_DR_def}) and include all the data given in Table \ref{quantities1}.  We use Monte Carlo simulation over the 14d parameter space and search for points $p=\left(p_1\ldots p_{14}\right)$ for which
\begin{equation}
\chi_\mathrm{I} (p) \le \chi_\mathrm{I}^{\rm exp},
\label{gfitI_cond}
\end{equation}
subject to the constraint
\begin{equation}
\Gamma[f_0(1370)\rightarrow \left(\pi\pi+ KK + \eta\eta\right)] < 500 \,\,{\rm MeV}.
\label{gfitI_const}
\end{equation}
In this case $\chi_\mathrm{I}^{\rm exp}= 7.3$. This leads to a set of points
\begin{equation}
S_\mathrm{I} = \left\{
 p \,| \, p\, \in \, \mathbb{R}^{14}: \, {\rm conditions\,\,(\ref{gfitI_cond})\,\, and \,\, (\ref{gfitI_const}) \,\, are \,\, upheld} \right\}.
\label{SI}
\end{equation}

\subsubsection{Global fit II}

In global fit II,   that we include the $\pi\pi$ and $KK$ partial decay widths  of $f_0(1370)$, the guiding function ($\chi_\mathrm{II}$) is computed from (\ref{chi_generic}) with all data in Table \ref{quantities1} included in $\chi_m$, $\chi_\Gamma$ and $\chi_{(\Gamma/\Gamma)}$.  For this case too, we use the same  iterative Monte Carlo method to search through the 14-dimensional parameter space for points at which
\begin{equation}
\chi_\mathrm{II}(p) \le \chi_\mathrm{II}^{\rm exp}.
\label{gfitII_cond}
\end{equation}
In this case $\chi_\mathrm{II}^{\rm exp}= 8.2$.  This leads to the second set
\begin{equation}
S_\mathrm{II} = \left\{
 p \,| \, p\, \in \, \mathbb{R}^{14}: \, {\rm condition \,\,(\ref{gfitII_cond})\,\, is \,\, upheld} \right\}.
 \label{SII}
\end{equation}

\subsubsection{Global fit III}

In global fit III,  target inputs are given in Table \ref{quantities2} in which  the decay ratios given by WA102 collaboration \cite{WA102} are not included.  Similar to the previous two global fits, we use the same iterative Monte Carlo method and scan the 14d parameter space for points at which
\begin{equation}
\chi_\mathrm{III}(p) \le \chi_\mathrm{III}^{\rm exp}.
\label{gfitIII_cond}
\end{equation}
where  $\chi_\mathrm{III}^{\rm exp}=5.6$.  This leads to the third set
\begin{equation}
S_\mathrm{III} = \left\{
 p \,| \, p\, \in \, \mathbb{R}^{14}: \, {\rm condition \,\,(\ref{gfitIII_cond})\,\, is \,\, upheld} \right\}.
 \label{SIII}
\end{equation}

For the three global fits I, II and III,   the results are  displayed in Fig.~\ref{F_gfits} and compared with their corresponding experimental data.   Our Monte Carlo simulations show that the points in the 14d parameter space (squares) that satisfy the global conditions (\ref{gfitI_cond}) and (\ref{gfitI_const}) for fit I,  condition (\ref{gfitII_cond}) for fit II, and condition (\ref{gfitIII_cond}) for fit III, lead to properties of $f_0(1370)$ that generally overlap with the experimental data (solid circles and error bars) with the exception of the input mass of $f_0(1370)$
that in our simulation comes out larger than its target experimental values displayed in Tables \ref{quantities1} and \ref{quantities2}.
We interpret this deviation of mass from its target values as a measure of the
size of the next  order corrections beyond the present leading order of the model  which falls in the range of $8\%-21\%$.    Also shown are the simulation averages and one standard deviation around the averages (triangles and error bars).

Fig.~\ref{F_gfits_comps} shows the quark and glue components of $f_0(1370)$ in the basis defined in (\ref{SNS_def}).    In this figure  the individual points (shown by ``+'') are just the results of the global fits without imposing any additional conditions.    Also shown are the averages (triangles) and  standard deviations (error bars).     The components are respectively proportional to ${\bar u}{\bar d} u d$, $({\bar d}{\bar s} d s + {\bar s}{\bar u} u s)
		/\sqrt{2}$, $s {\bar s}$,
		$(u {\bar u}  + d {\bar d})/\sqrt{2}$ and
		glue $G$.   For convenience, the total percentages of the four-quark, the quark-antiquark and the glue  for each case are also given beneath their detailed component figure in Fig. \ref{F_gfits_comps}.
Clearly,  the global fits show that the four-quark and glue components are considerably smaller than the quark-antiquark  components.       The $s {\bar s}$ is the dominant component in either case.  We will of course further zoom in on the properties of $f_0(1370)$ in the next subsection, however, we will see that the conclusion will remain unchanged.   For the convenience of the reader, we have also made histograms in Fig. \ref{F_gfits_hist} that show the distribution of the simulations for each component by breaking them down into five bins.   This makes it easier to see that the two quark components (particularly {$s {\bar s}$) are the only ones that have high-percentage bins filled.   The substructure of $f_0(1370)$ shown for the global fits I, II and III in Figs. \ref{F_gfits_comps} and \ref{F_gfits_hist}  has been a consistent and stable trend in all our simulations, regardless of the additional conditions or filters imposed.       This is somewhat in contrast with the case of other isosinglets studied in this model [$f_0(500)$, $f_0(980)$, $f_0(1500)$ and $f_0(1710)$] for which their substructure are rather sensitive and closely correlated (see \cite{FAA_1500}).
\begin{figure}[h]
	\begin{center}
		\epsfxsize = 5.5 cm
		 \epsfbox{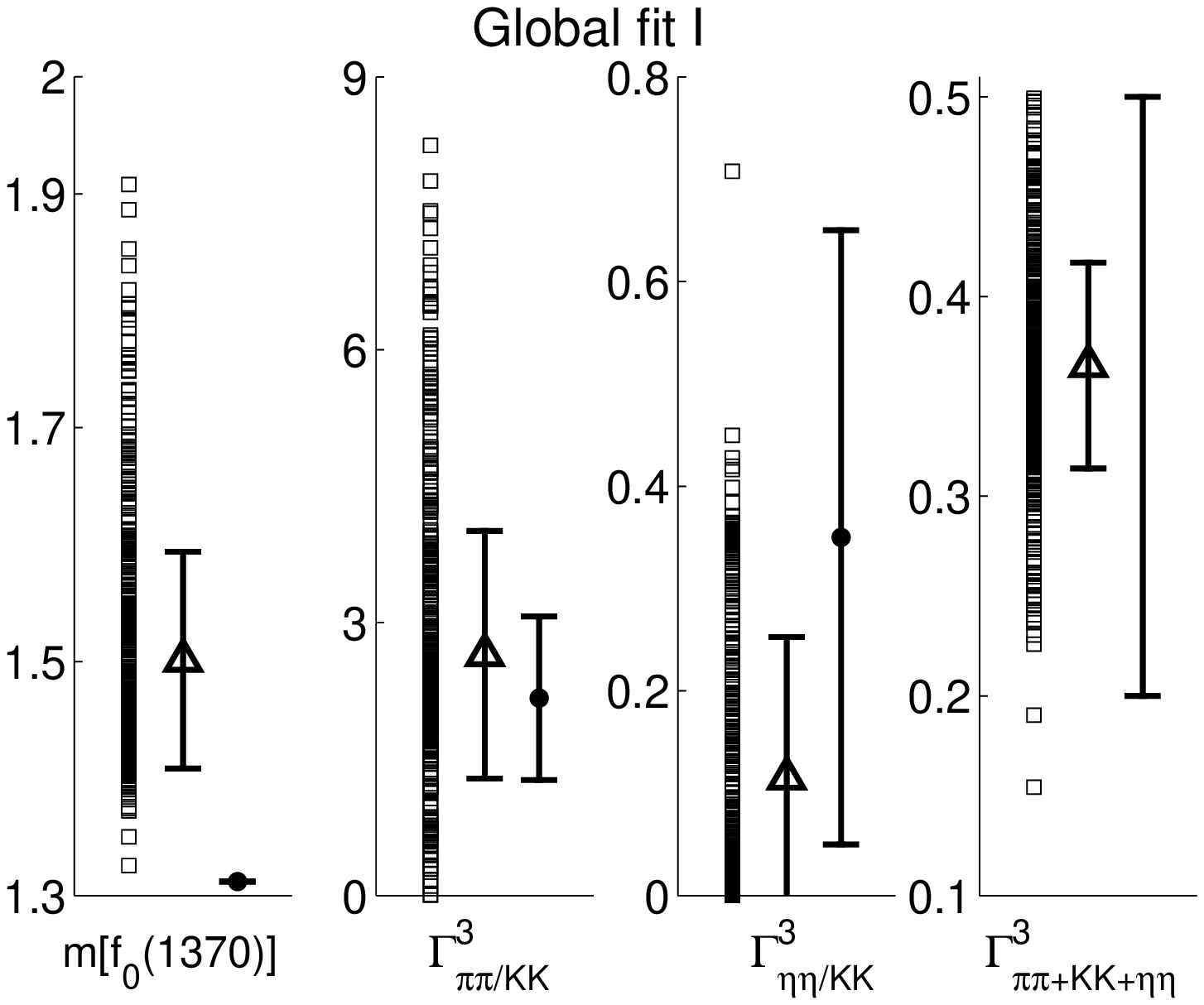}
		 \hskip .7cm
		\epsfxsize = 5.5 cm
		 \epsfbox{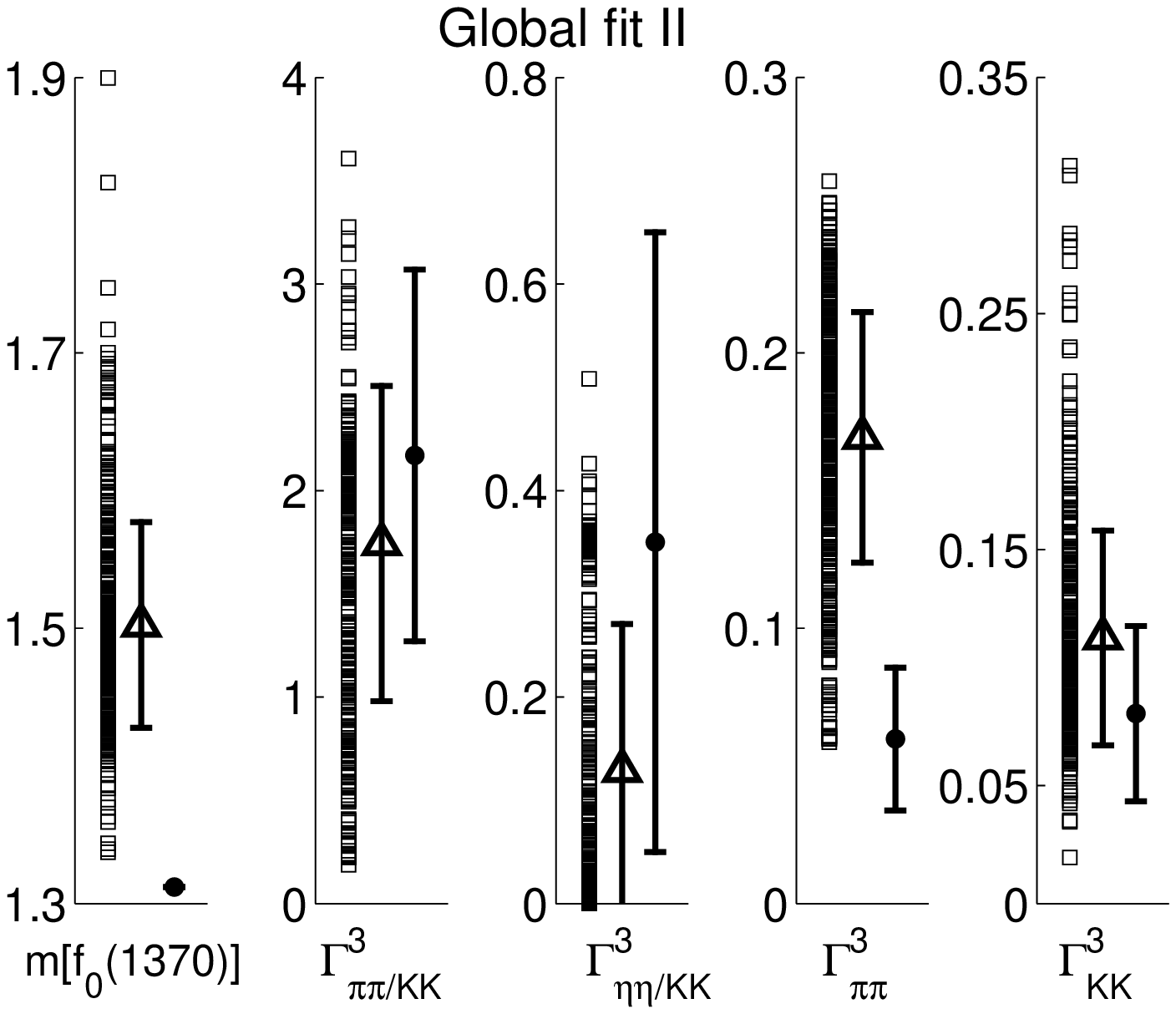}
		 \hskip .5 cm
		\epsfxsize = 5.5 cm
		 \epsfbox{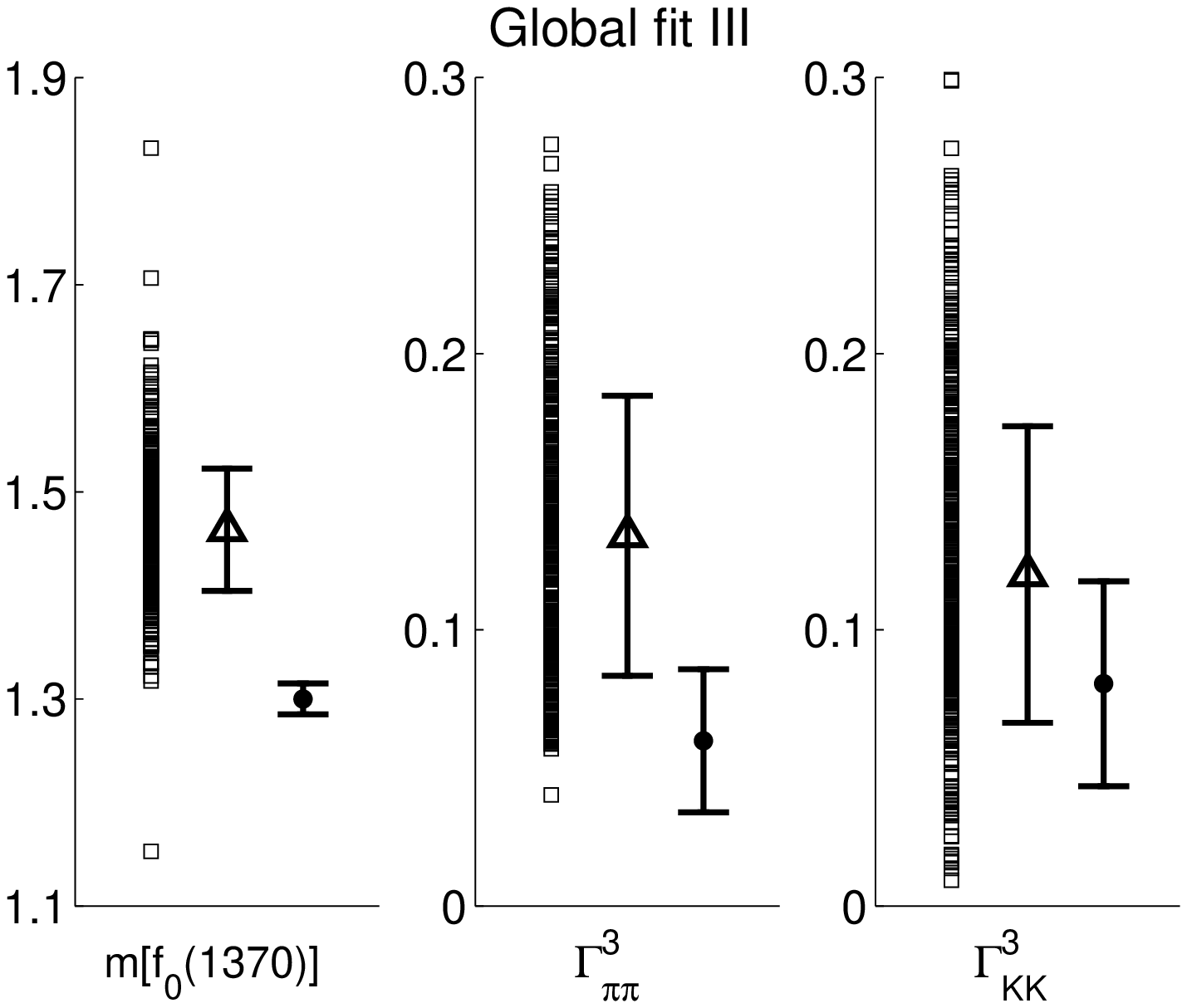}
		\caption{Mass and decay properties (GeV) of $f_0(1370)$ obtained from Monte Carlo simulation for global fits I (left),  II (middle) and III (right)  defined in subsection IV-A. The results of simulations (squares) and their averages and one standard deviation around the averages (triangles and error bars)  are compared with experiment (filled circles and error bars). }
		\label{F_gfits}
	\end{center}
\end{figure}
\begin{figure}[h]
	\begin{center}
		\epsfxsize = 5.5 cm
		 \epsfbox{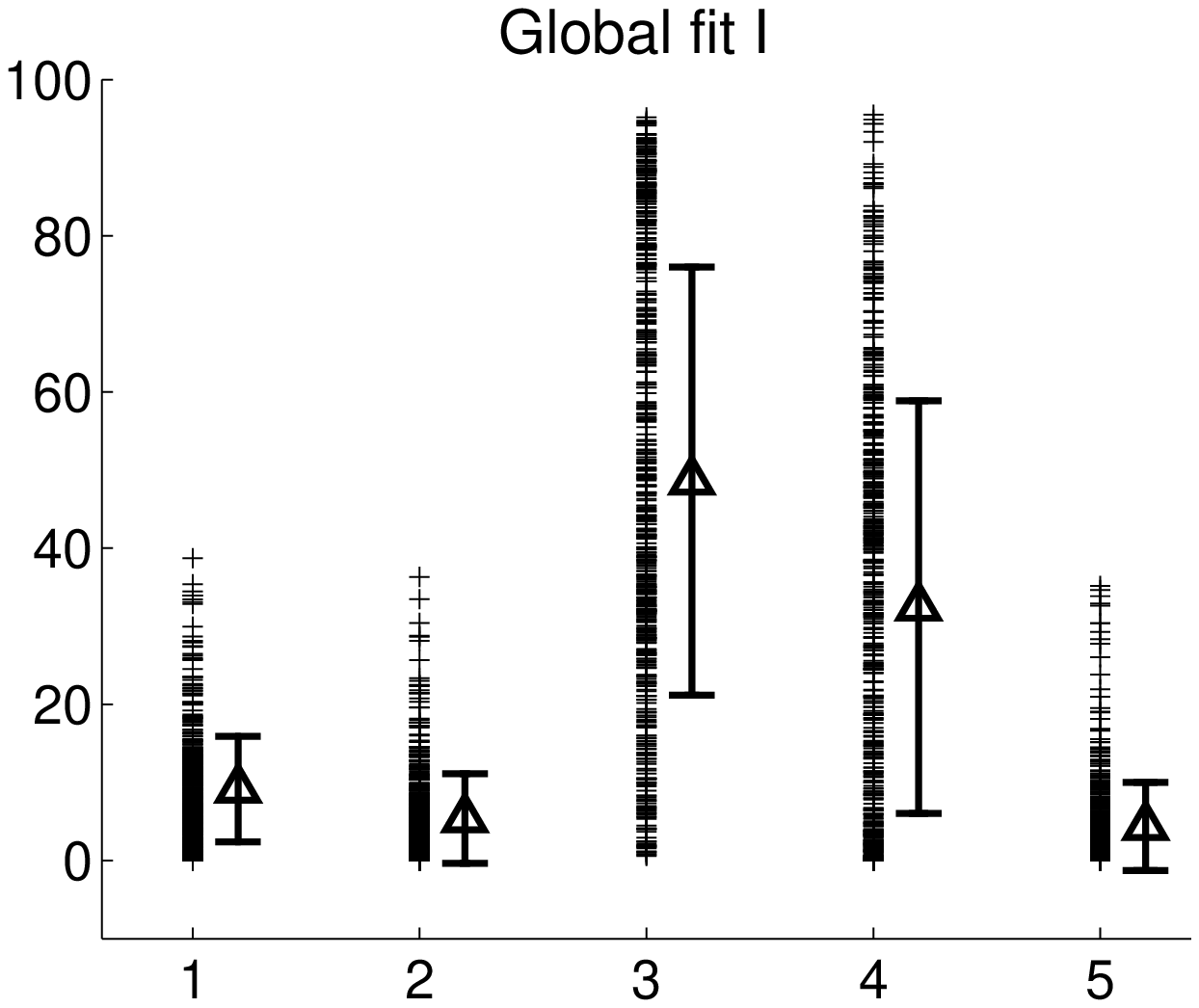}
		 \hskip .5cm
		\epsfxsize = 5.5 cm
		 \epsfbox{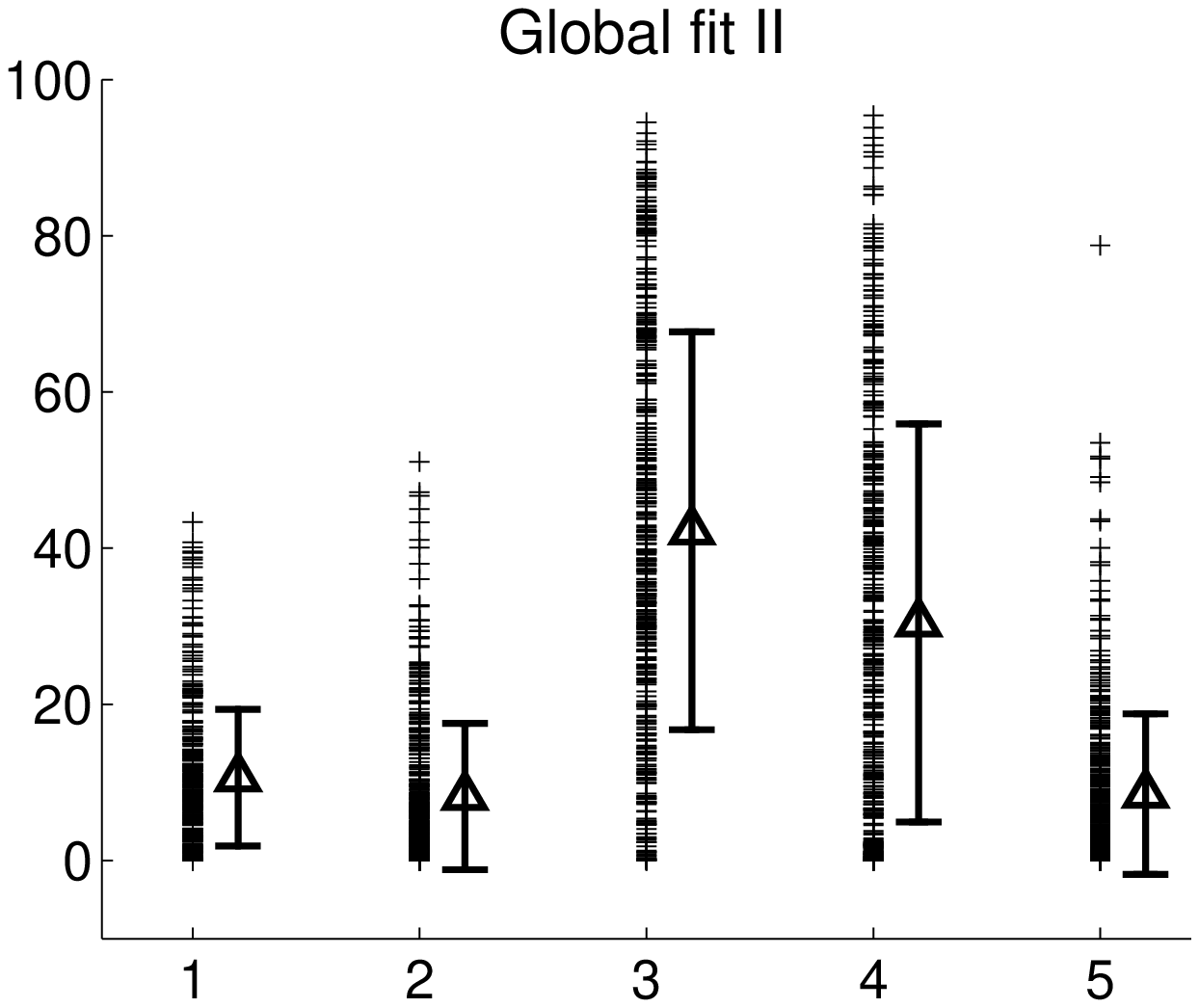}
		 \hskip .5cm
		 \epsfxsize = 5.5 cm
	 	 \epsfbox{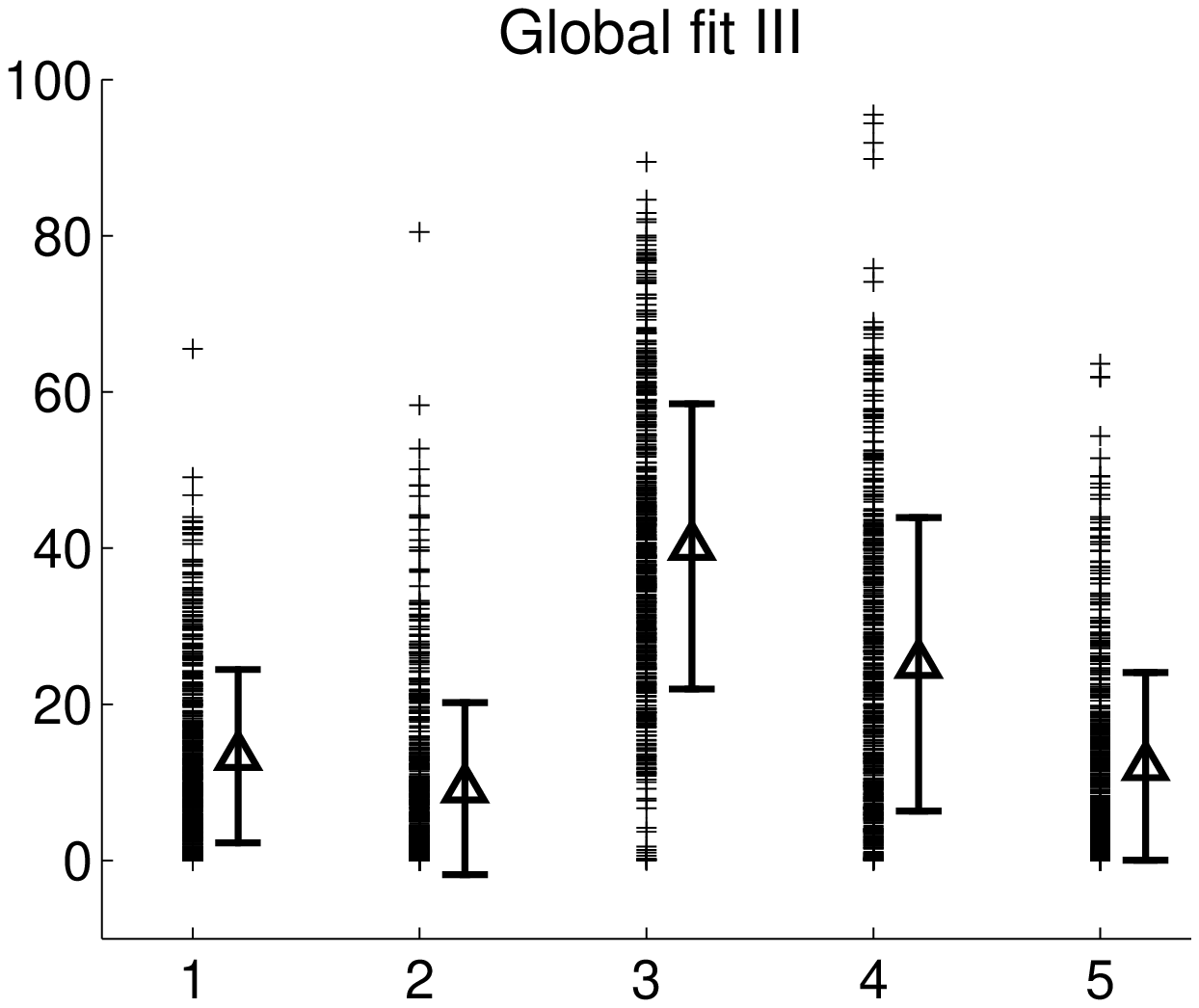}
		\epsfxsize = 5.5 cm
		 \epsfbox{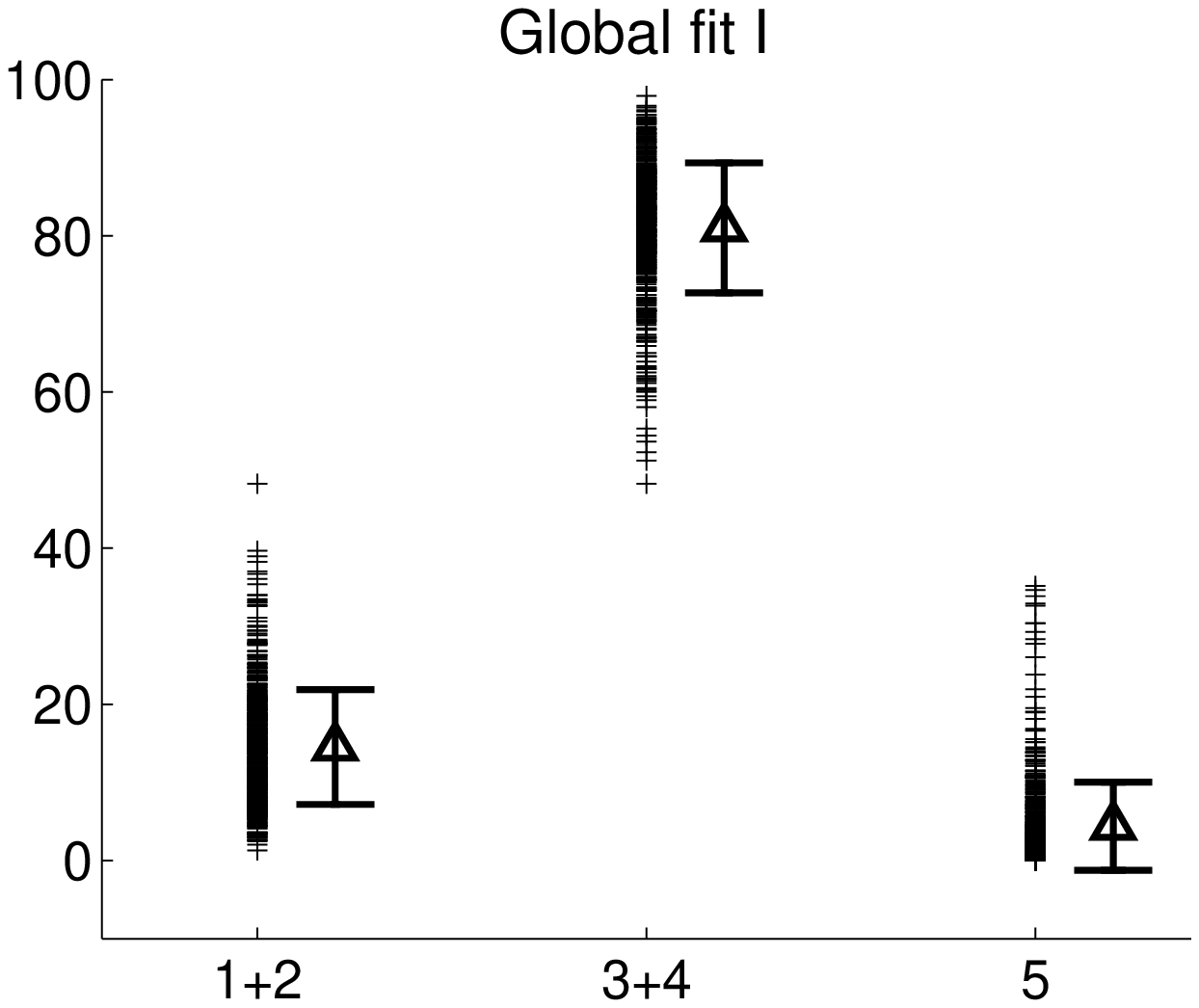}
		 \hskip .5cm
		\epsfxsize = 5.5 cm
		 \epsfbox{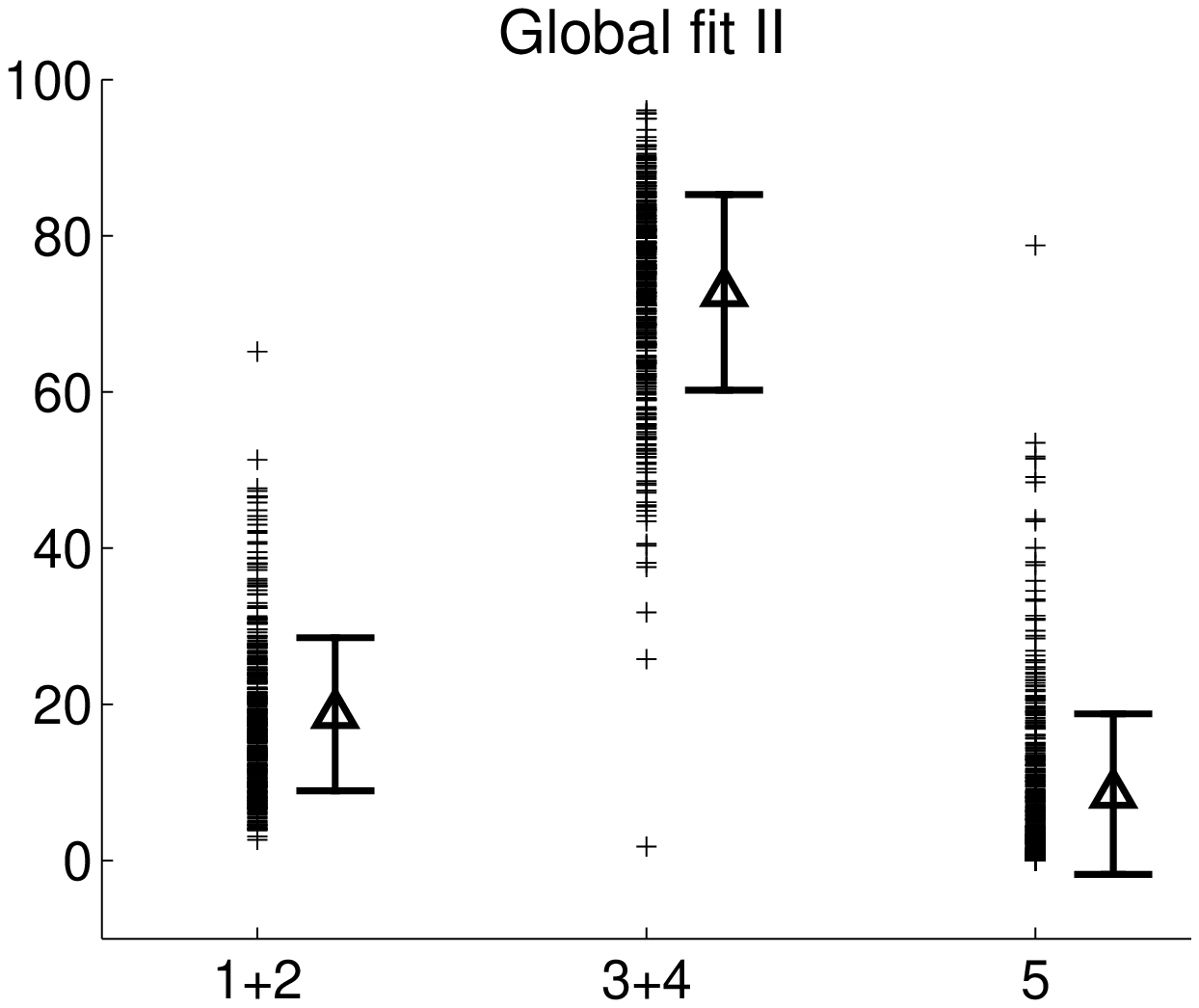}
     	 \hskip .5cm
		\epsfxsize = 5.5 cm
	 	 \epsfbox{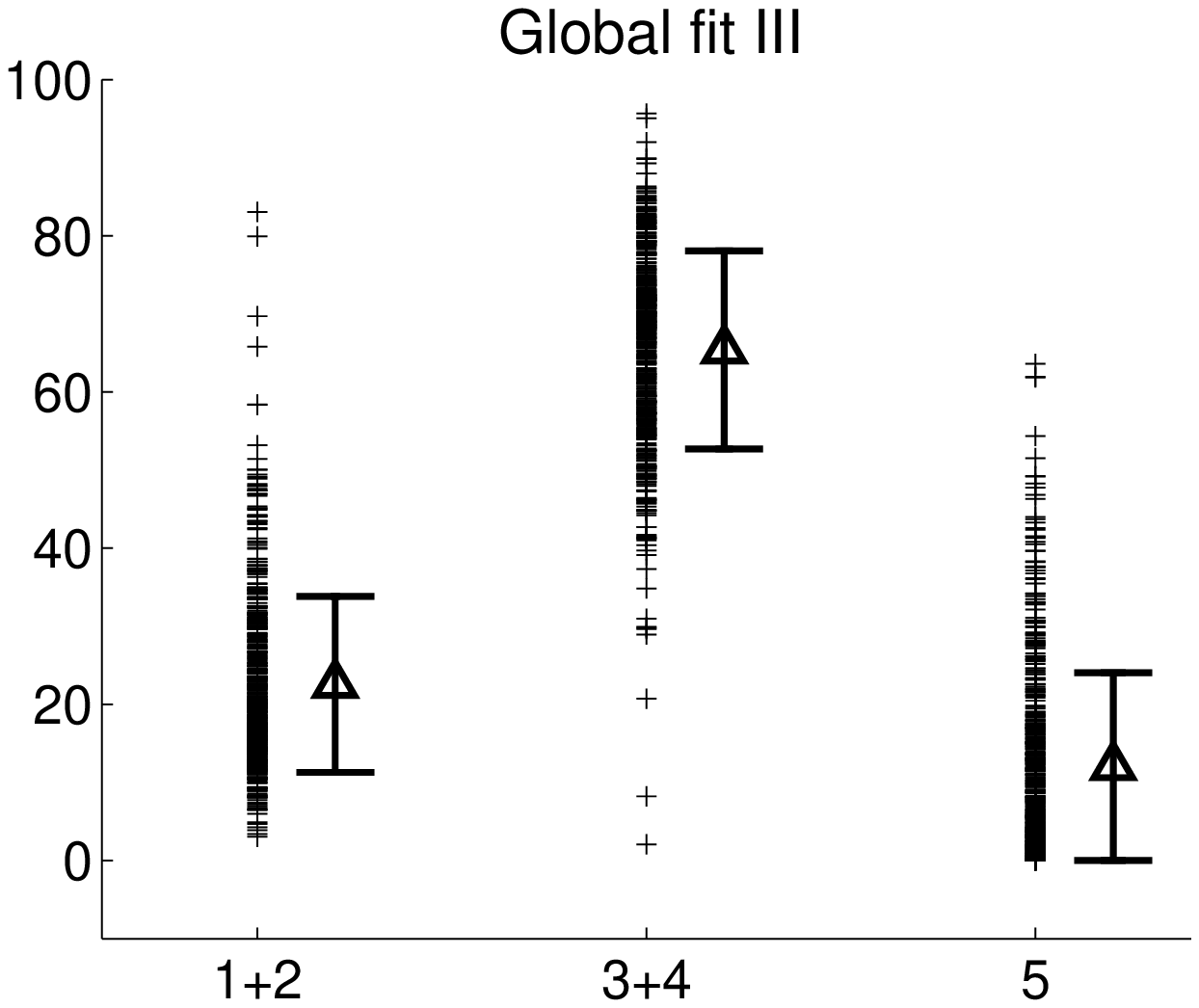}
		\caption{Percentages of quark and glue components of $f_0(1370)$ obtained from Monte Carlo simulation for global fits I (left), II (middle) and III (right) defined in subsection IV-A.  In the first row, the components 1 to 5 on the horizontal axes respectively represent
		${\bar u}{\bar d} u d$, $( {\bar d}{\bar s} d s + {\bar s}{\bar u} s u)
		/\sqrt{2}$, $s {\bar s}$,
		$(u {\bar u}  + d {\bar d})/\sqrt{2}$, and
		glueball $G$. In the second row, the percentages of total four-quark (1+2), total quark-antiquark (3+4) and glue (5) are given. }
		\label{F_gfits_comps}
	\end{center}
\end{figure}
\begin{figure}[h]
	\begin{center}
		\vskip .75cm
		\epsfxsize = 5.5 cm
		 \epsfbox{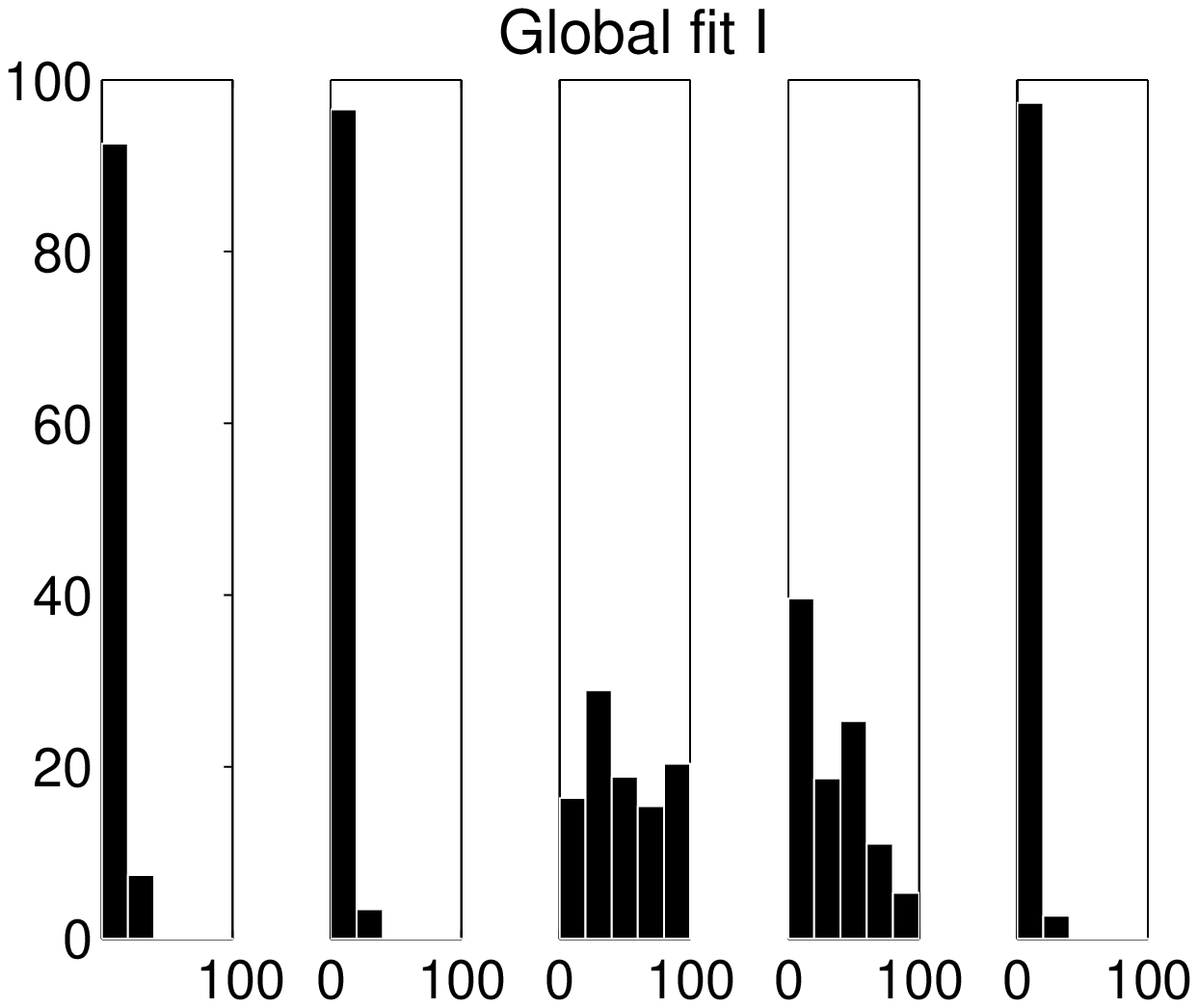}
		 \hskip .5 cm
		\epsfxsize = 5.5 cm
		 \epsfbox{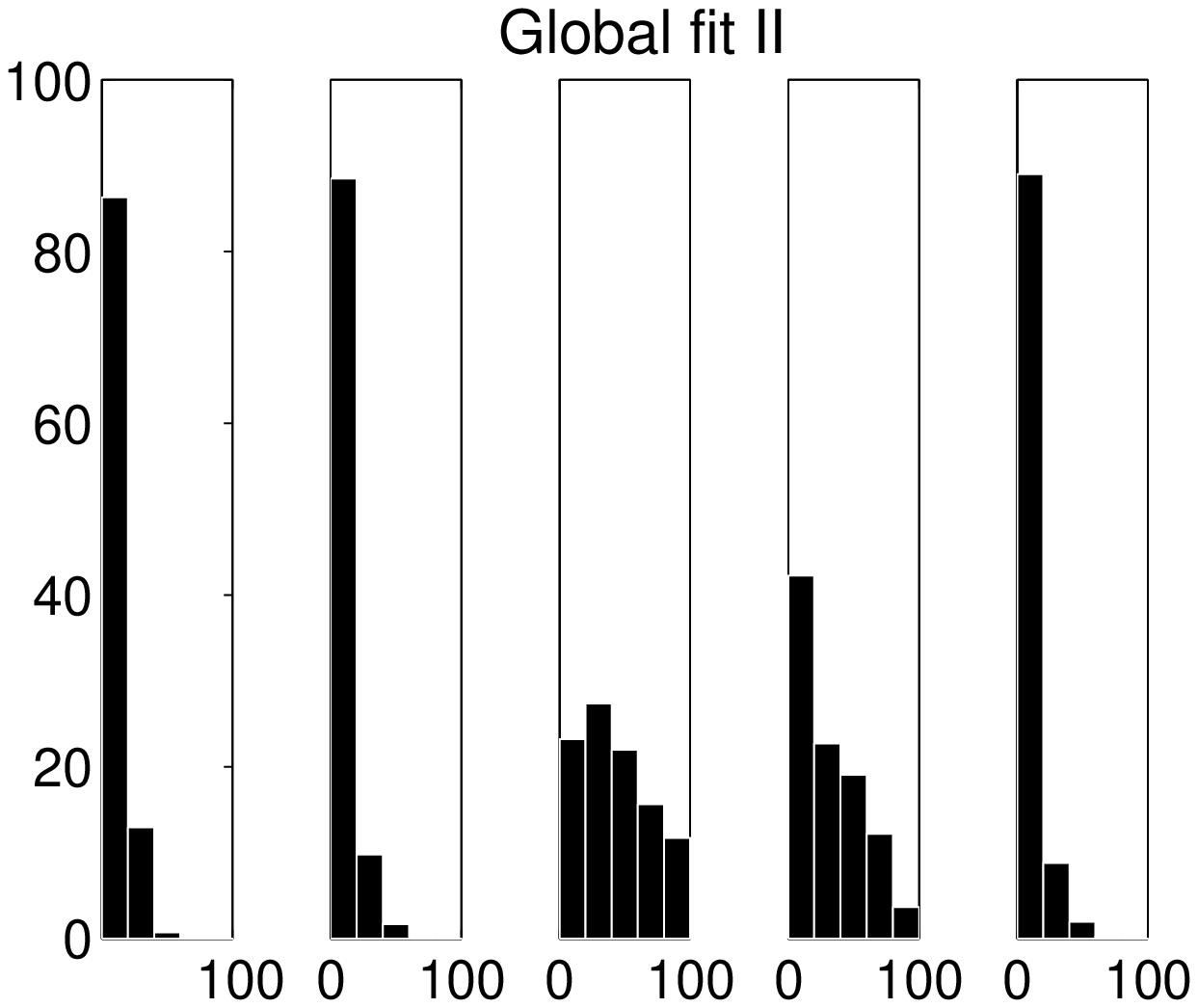}
	     \hskip .5 cm
	     \epsfxsize = 5.5 cm
		 \epsfbox{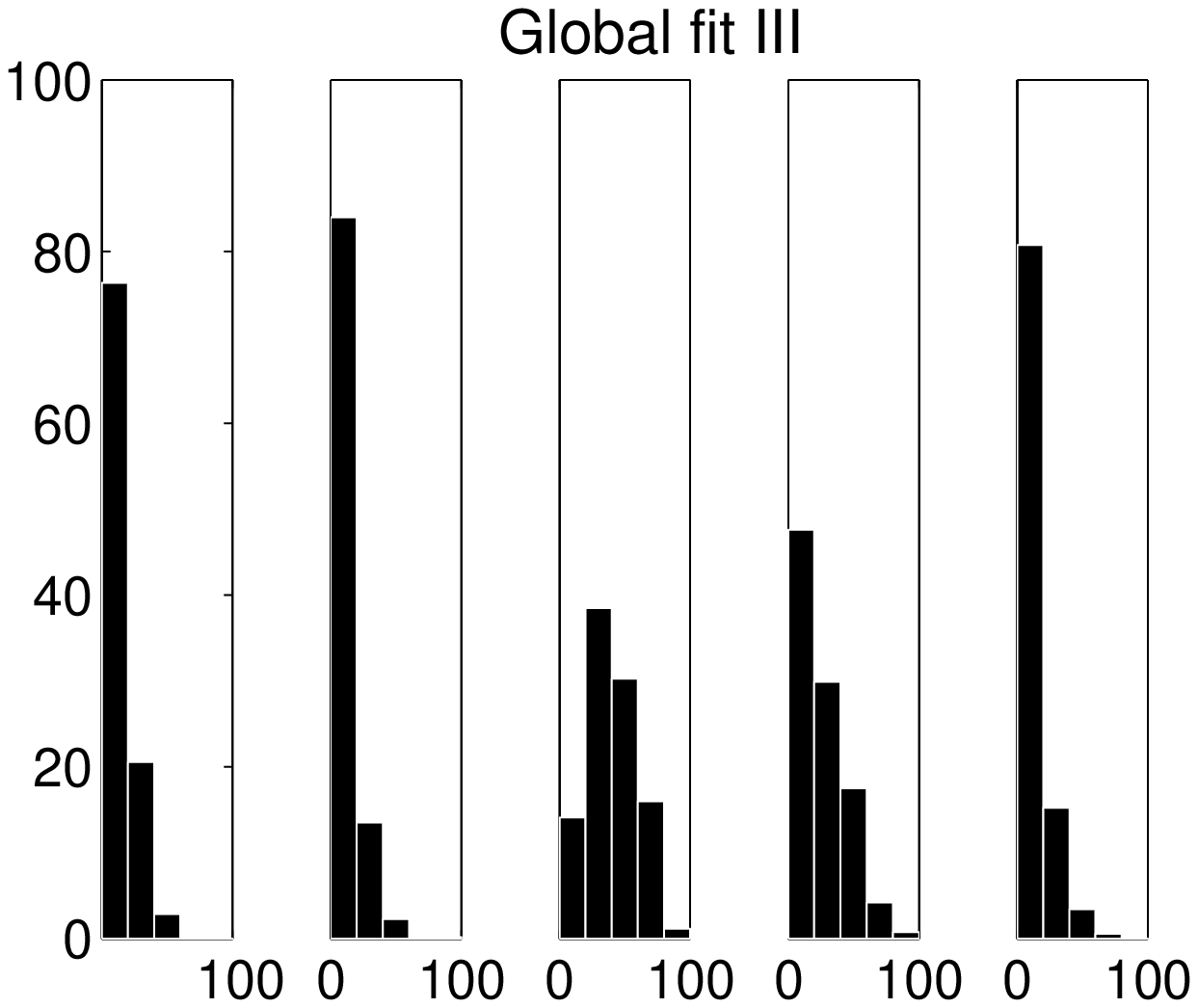}
		\caption{
		Histograms of Monte Carlo simulations versus each component of $f_0(1370)$ obtained
		in global fits I (left set of figures),  II (middle  set of figures) and III (right set of figures).  Each set consists of five figures,
the vertical axes represent the percentages of Monte Carlo simulations, the horizontal axes (in each set)
from left to right respectively represent the percentages of
		components
		${\bar u}{\bar d} u d$, $({\bar d}{\bar s} d s + {\bar s}{\bar u} s u)
		/\sqrt{2}$, $s {\bar s}$,
		$(u {\bar u}  + d {\bar d} )/\sqrt{2}$, and
		glueball $G$.
		Each component is divided into five percentage intervals (0-20\%, 20\%-40\%, etc).
        For example, the first figure on the left shows that more than 90\% of Monte Carlo simulations for global fit I, resulted in estimating the ${\bar u}{\bar d} u d$ component of $f_0(1370)$ to be below 20\%, and less than 10\% of the simulations resulted in estimating the  $({\bar d}{\bar s} d s + {\bar s}{\bar u} s u)
		/\sqrt{2}$ component between 20\%$-$40\%, etc.
		      }
		\label{F_gfits_hist}
	\end{center}
\end{figure}
\subsection{Zooming in on $f_0(1370)$}

In previous subsection we made three global fits to the mass spectrum and decay properties of isosinglet scalar states below and above 1 GeV.    This led to identification of three sets of points in the 14d parameter space [$S_\mathrm{I}$, $S_\mathrm{II}$ and $S_\mathrm{III}$ defined in (\ref{SI}), (\ref{SII}) and (\ref{SIII}),  respectively] for which there is an overall agreement between the model predictions and their corresponding experimental data.  Working within these three sets  ensures that the model respects (at least qualitatively) the global properties of all of these isosinglet scalar states.    In this subsection,  we  further refine our investigation of $f_0(1370)$ by searching for subsets within each of these three sets that better describe this state.

\subsubsection{Zoom A}

First,  we search through set $S_\mathrm{I}$ for a subset that gives an overall agreement between the model predictions for properties of $f_0(1370)$ and their corresponding experimental data by imposing
\begin{equation}
\chi_\mathrm{I}^{1370} \le \left(\chi_\mathrm{I}^{1370}\right)^{\rm exp}
\label{SIA_def}
\end{equation}
with
\begin{equation}
\chi_\mathrm{I}^{1370} = \chi^{1370}_m + \chi^{1370}_{(\Gamma/\Gamma)},
\end{equation}
where $\left(\chi_\mathrm{I}^{1370}\right)^{\rm exp}=1.3$.   This leads to a subset $S_\mathrm{IA} \subset S_\mathrm{I}$ defined by
\begin{equation}
S_\mathrm{IA} = \left\{
 p \,| \, p \in S_\mathrm{I} \,:\, {\rm relation\,\,(\ref{SIA_def})\,\, is \,\, upheld} \right\}.
 \label{SIA}
\end{equation}
The results for the components of $f_0(1370)$ over subset $S_\mathrm{IA}$ are shown in Fig.~\ref{F_comps_zoom1}
(left panel) where the components at a given point in this subset are shown by a ``$+$'' together with the averages (triangles) and the standard deviations around the averages (error bars). We see that while zooming in on $f_0(1370)$ slightly shifts the components compared to the global fit I, the general structure remains the same, namely that  the quark-antiquark components (particularly $s{\bar s}$) remain dominant.

Similarly, we search through set $S_\mathrm{II}$ for a subset that better describes the overall properties of $f_0(1370)$ by imposing the condition
\begin{equation}
\chi_\mathrm{II}^{1370} \le \left(\chi_\mathrm{II}^{1370}\right)^{\rm exp}
\label{chiII}
\end{equation}
with
\begin{equation}
\chi_\mathrm{II}^{1370} = \chi^{1370}_m + \chi^{1370}_\Gamma + \chi^{1370}_{(\Gamma/\Gamma)},
\end{equation}
where $\left(\chi_\mathrm{II}^{1370}\right)^{\rm exp}$=2.2.   This leads to a subset $S_\mathrm{IIA} \subset S_\mathrm{II}$
\begin{equation}
S_\mathrm{IIA} = \left\{
 p \,| \, p \in S_\mathrm{II} \,:\, {\rm relation\,\,(\ref{chiII})\,\, is \,\, upheld} \right\}.
 \label{SIIA}
\end{equation}
The results for the components are shown in Fig.~\ref{F_comps_zoom1} (middle panel) which shows the same characteristics  as those in Fig.~\ref{F_gfits_comps} in which the quark-antiquark components (specially $s{\bar s}$) dominate.

Finally, we search through set $S_\mathrm{III}$ for a subset in which the properties of $f_0(1370)$ are better described.  We  impose the condition
\begin{equation}
\chi_\mathrm{III}^{1370} \le \left(\chi_\mathrm{III}^{1370}\right)^{\rm exp}
\label{chiIII}
\end{equation}
with
\begin{equation}
\chi_\mathrm{III}^{1370} = \chi^{1370}_m + \chi^{1370}_\Gamma, 
\end{equation}
where $\left(\chi_\mathrm{III}^{1370}\right)^{\rm exp}=0.9$.  This leads to the subset $S_\mathrm{IIIA} \subset S_\mathrm{III}$
\begin{equation}
S_\mathrm{IIIA} = \left\{
 p \,| \, p \in S_\mathrm{III} \,:\, {\rm relation\,\,(\ref{chiIII})\,\, is \,\, upheld} \right\}.
 \label{SIIIA}
\end{equation}
The results for the components are shown in Fig.~\ref{F_comps_zoom1} (right panel) which again shows a similar behavior   as those in Fig.~\ref{F_gfits_comps} where the quark-antiquark components (specially $s{\bar s}$) dominate.

This figure shows the robustness of the results and that when zooming in on $f_0(1370)$, while we get subsets that better describe the overall properties of this state,  its components do not change much.    As stated previously, this is not the case with other isosinglet states and in that sense $f_0(1370)$ is evidently a special case.    Simulation histograms versus components are displayed in Fig. \ref{F_hist_zoom} showing the preeminence of quark-antiquark components in $f_0(1370)$.

\subsubsection{Zoom B}

We can further zoom in by applying a more stringent condition on each set and examine whether the components of $f_0(1370)$ retain their pattern observed above.     For this purpose we search for a subset $S_\mathrm{IB}\in S_\mathrm{I}$ such that
all  decay ratios that were used as part of defining $S_\mathrm{I}$) are within their expected ranges (since the mass of $f_0(1370)$  has been fixed by WA102 collaboration to 1312 MeV, we do not impose that highly constraining codition in this zoom). This means that for points in $S_\mathrm{IB}$ the decay ratios of $f_0(1370)$ should fall within their experimental values,
\begin{equation}
S_\mathrm{IB} = \left\{
 p \,| \, p \in S_\mathrm{I} \,:\,  \Gamma^3_{\pi\pi/KK}, \Gamma^3_{\eta\eta/KK} {\rm \,\, are \,\, within \,\, their \,\, experimental \,\, ranges} \right\}.
 \label{SIB}
\end{equation}
The results for the components are shown in Fig.~\ref{F_comps_zoom2}, where  again we see that the $s{\bar s}$ component is quite pronounced in agreement with the preceding discussions.    We have given in Fig. \ref{F_ss_vs_nn} a comparison of the $s{\bar s}$ versus $n{\bar n}$ components for the three global fits and the zooms discussed in this section.   We see that for the majority of the simulations the $s{\bar s}$ component dominates.

We impose similar strong constrains on the other global sets II and III but we do not find any subsets where all the inputs for $f_0(1370)$ can be met.

\begin{figure}[h]
\begin{center}
\vskip .75cm
\epsfxsize = 5.5 cm
 \epsfbox{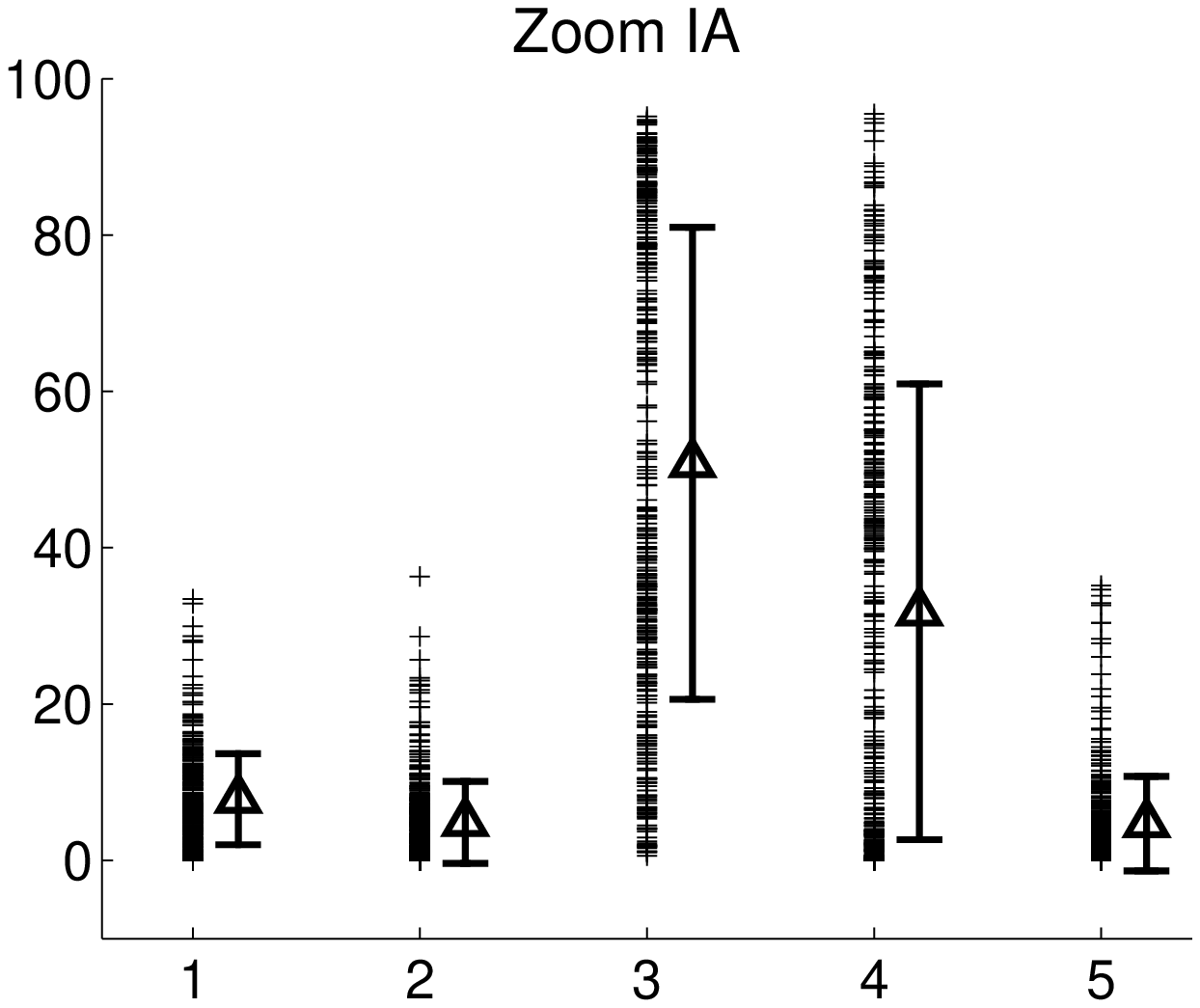}
 \hskip .5 cm
\epsfxsize = 5.5 cm
 \epsfbox{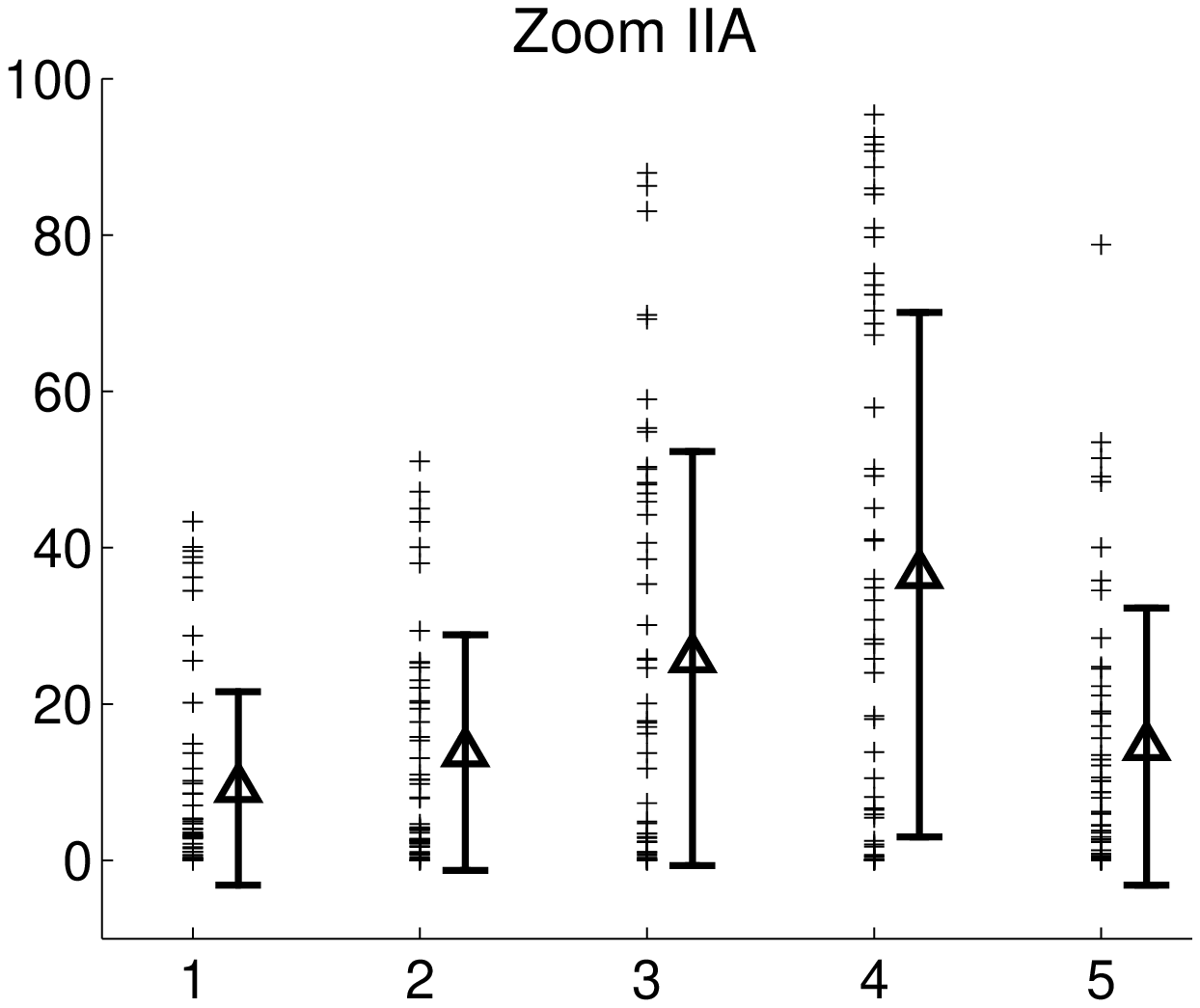}
\hskip .5cm
\epsfxsize = 5.5 cm
 \epsfbox{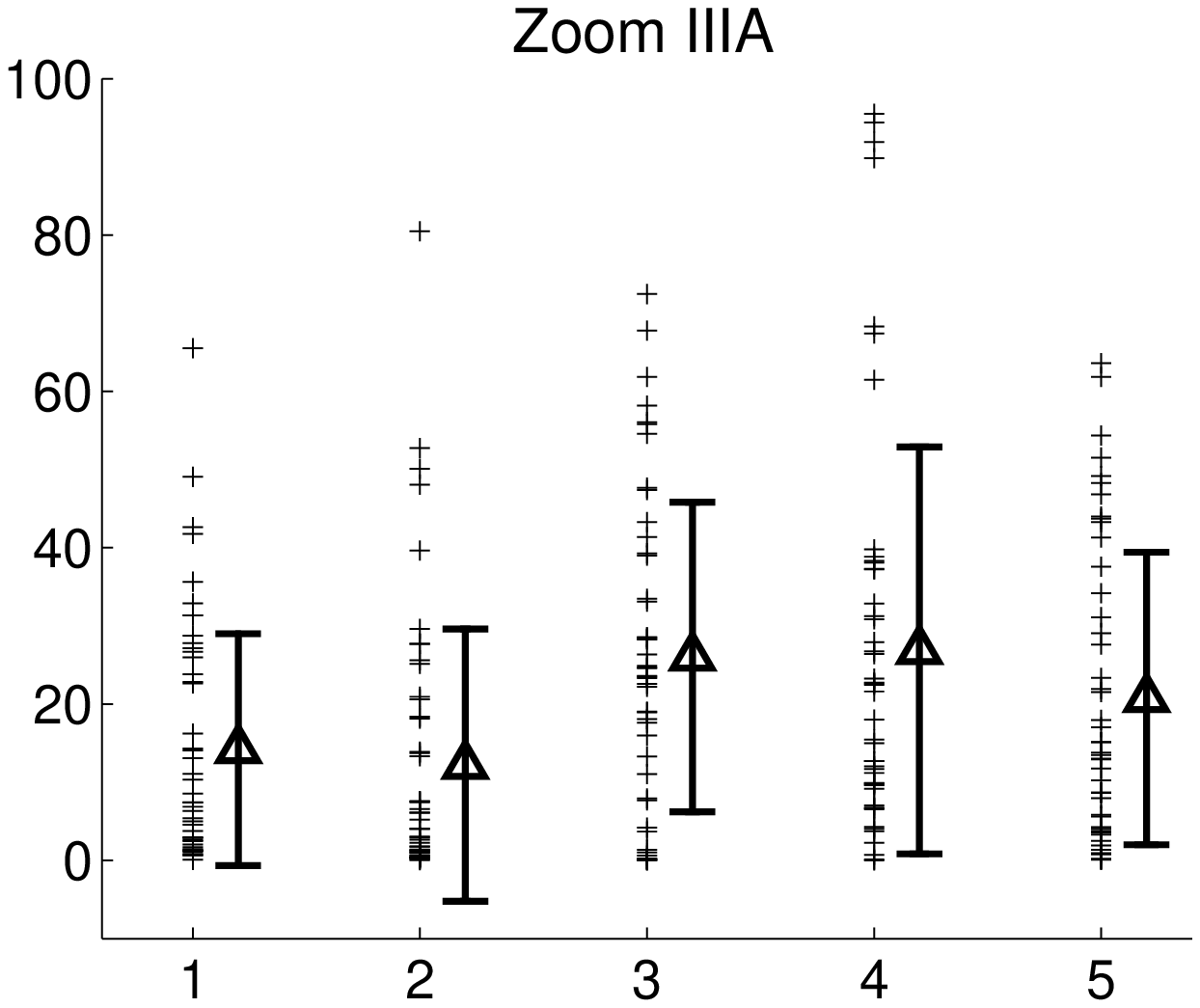}
\epsfxsize = 5.5 cm
 \epsfbox{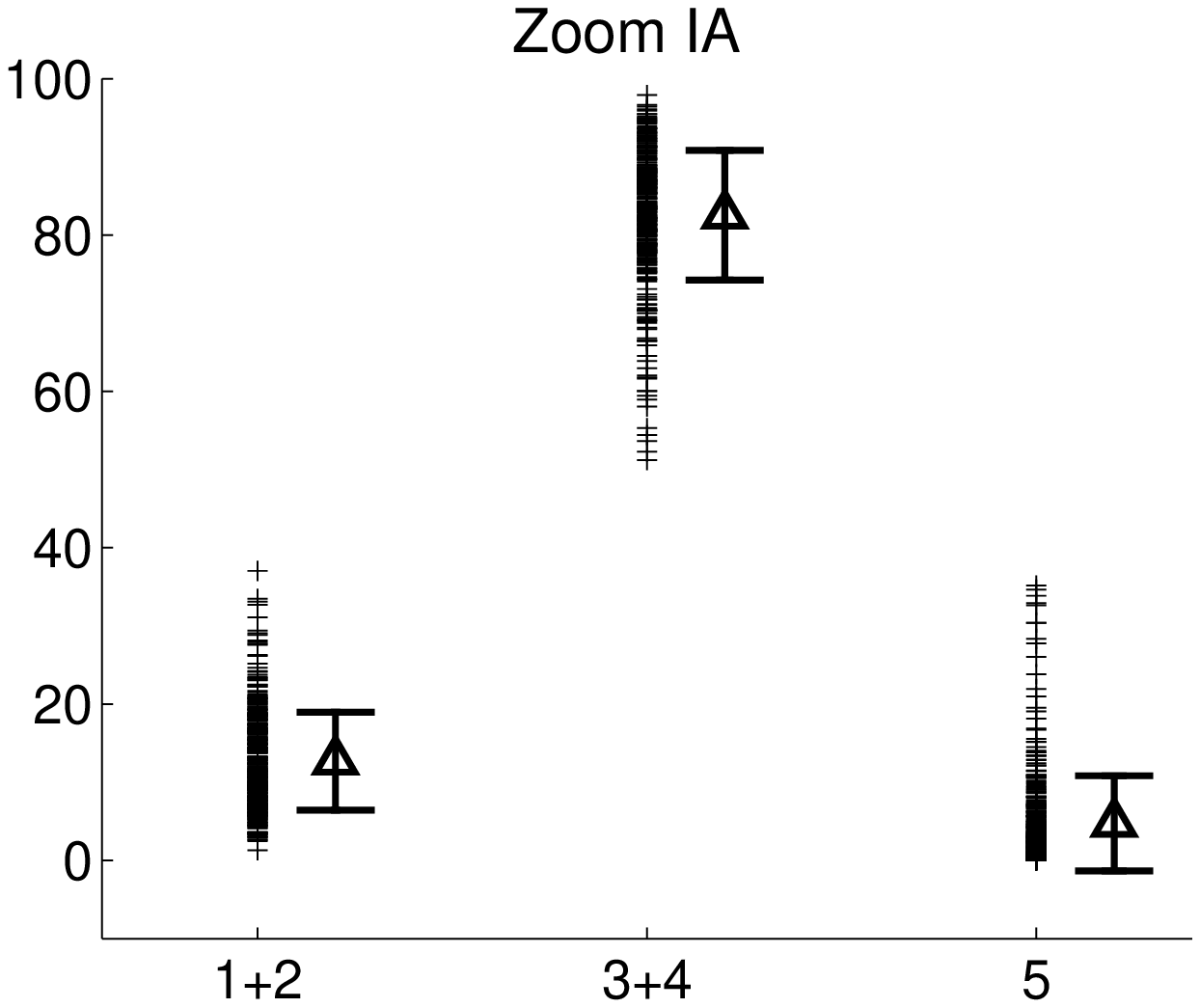}
\hskip .5cm
\epsfxsize = 5.5 cm
 \epsfbox{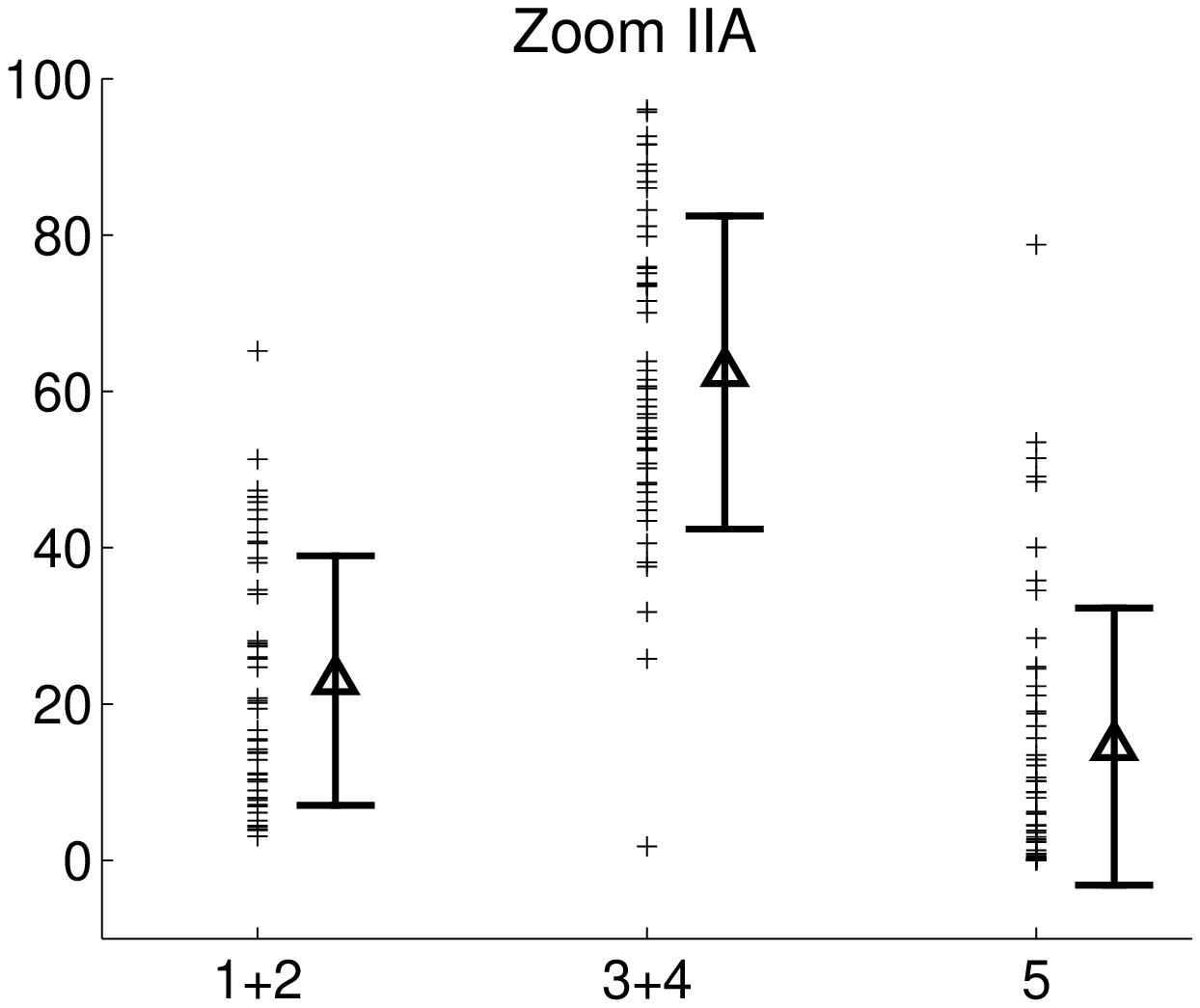}
 \hskip .5 cm
\epsfxsize = 5.5 cm
 \epsfbox{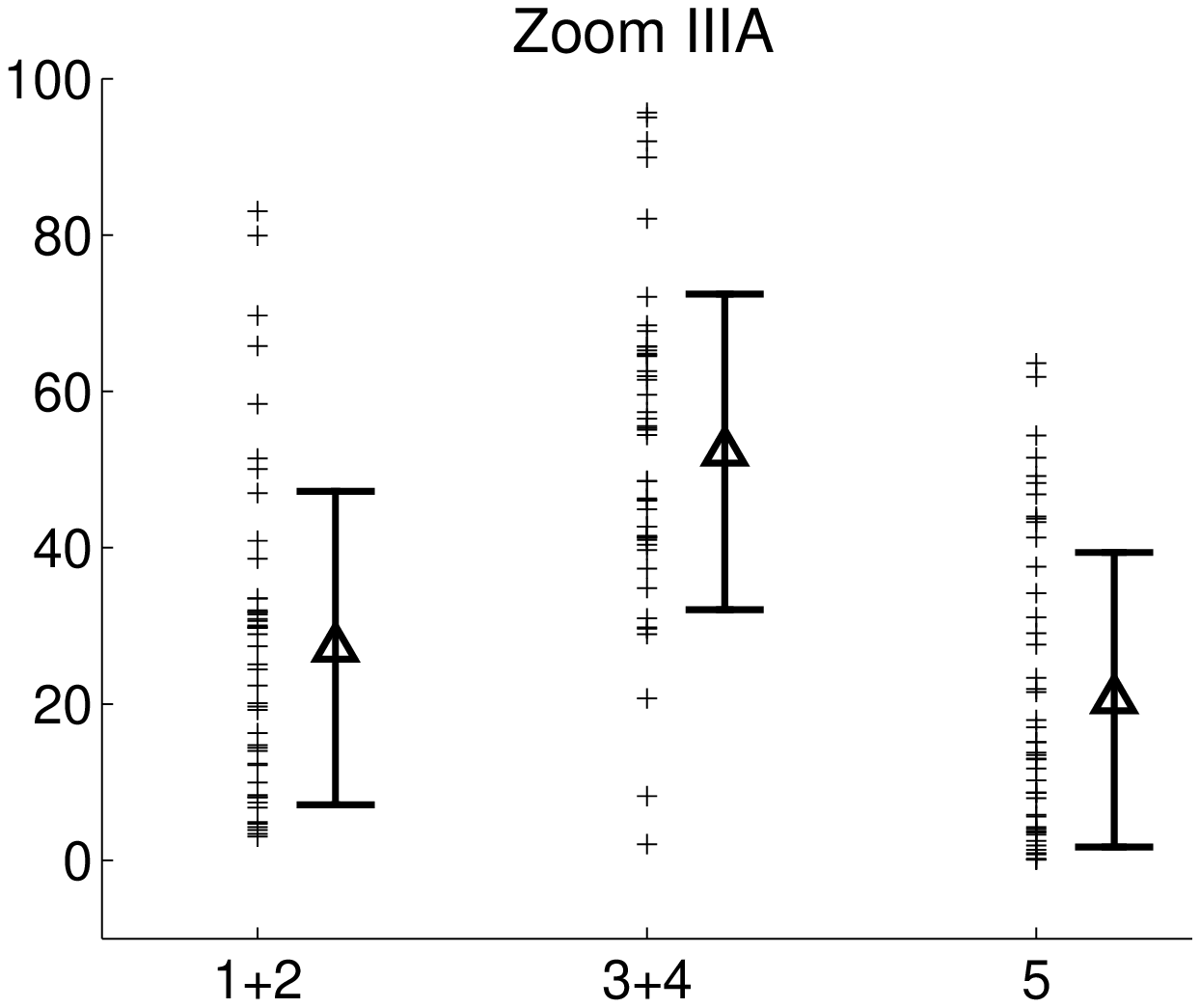}
\caption{Percentages of the quark and glue components of $f_0(1370)$ obtained obtained
over subsets $S_\mathrm{IA}$ [relation (\ref{SIA})],   $S_\mathrm{IIA}$ [relation (\ref{SIIA})] and $S_\mathrm{III}$ [relation (\ref{SIIIA}].
  Components 1 to 5 respectively represent
		${\bar u}{\bar d} u d$, $( {\bar d}{\bar s} d s + {\bar s}{\bar u} s u)
		/\sqrt{2}$, $s {\bar s}$,
		$(u {\bar u}  + d {\bar d})/\sqrt{2}$, and
		glue.
In the second row, the percentages of total four-quark (1+2), total quark-antiquark (3+4) and glue (5) are given.
The averages (triangles) and standard deviations (error bars) are shown.
}
\label{F_comps_zoom1}
\end{center}
\end{figure}

\begin{figure}[h]
\begin{center}
\vskip .75cm
\epsfxsize = 5.5 cm
 \epsfbox{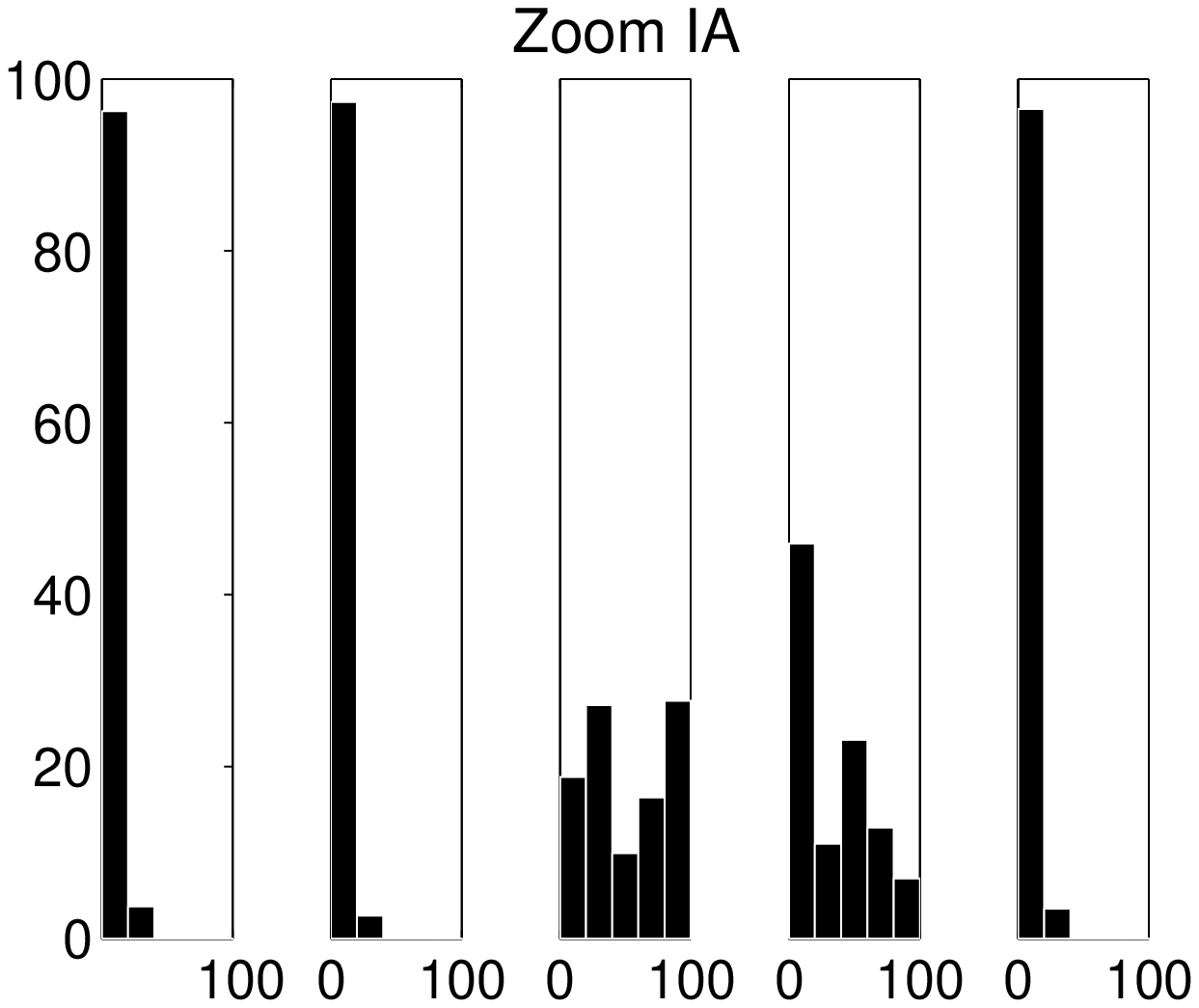}
 \hskip .5 cm
\epsfxsize = 5.5 cm
 \epsfbox{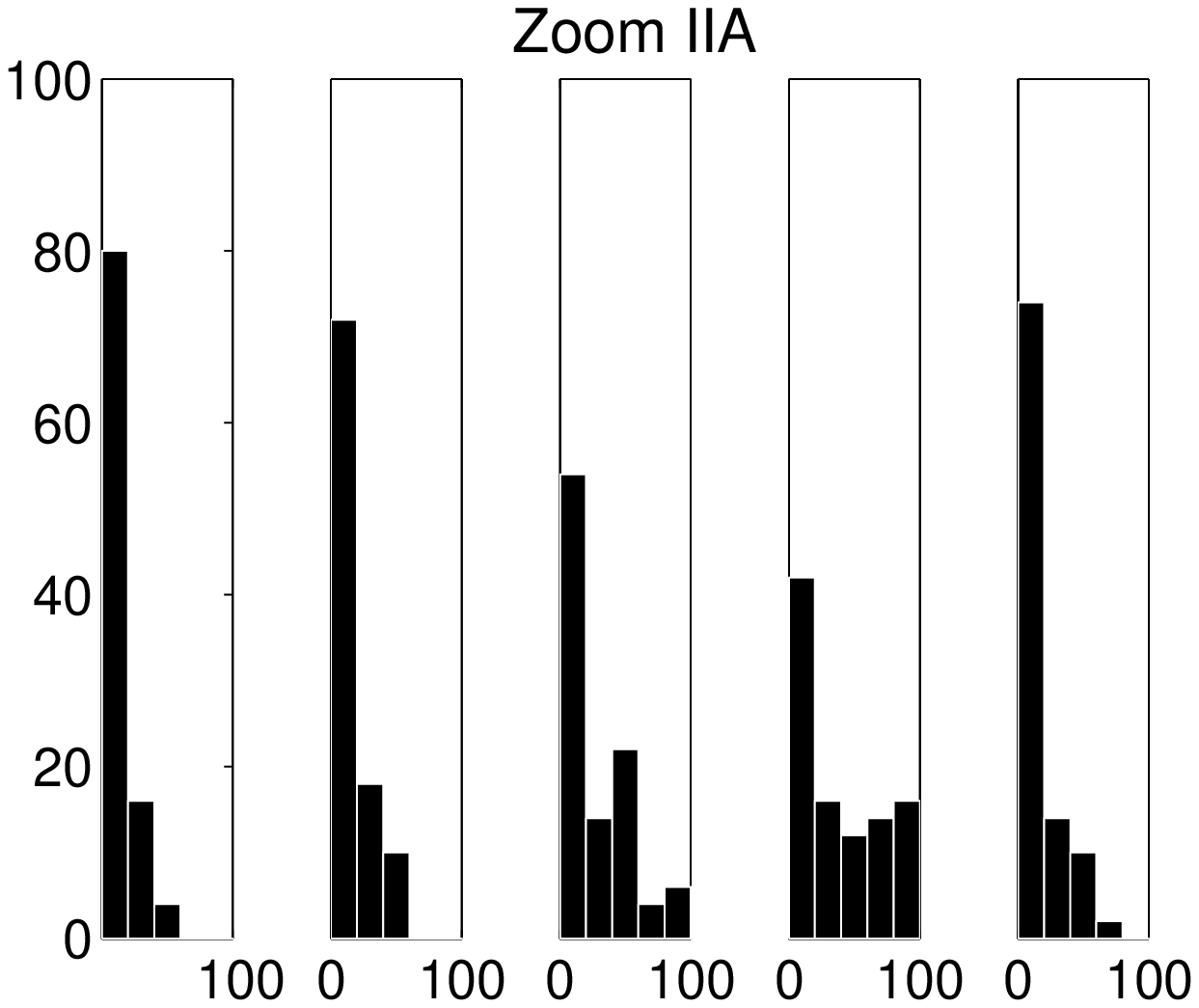}
 \hskip .5 cm
\epsfxsize = 5.5 cm
 \epsfbox{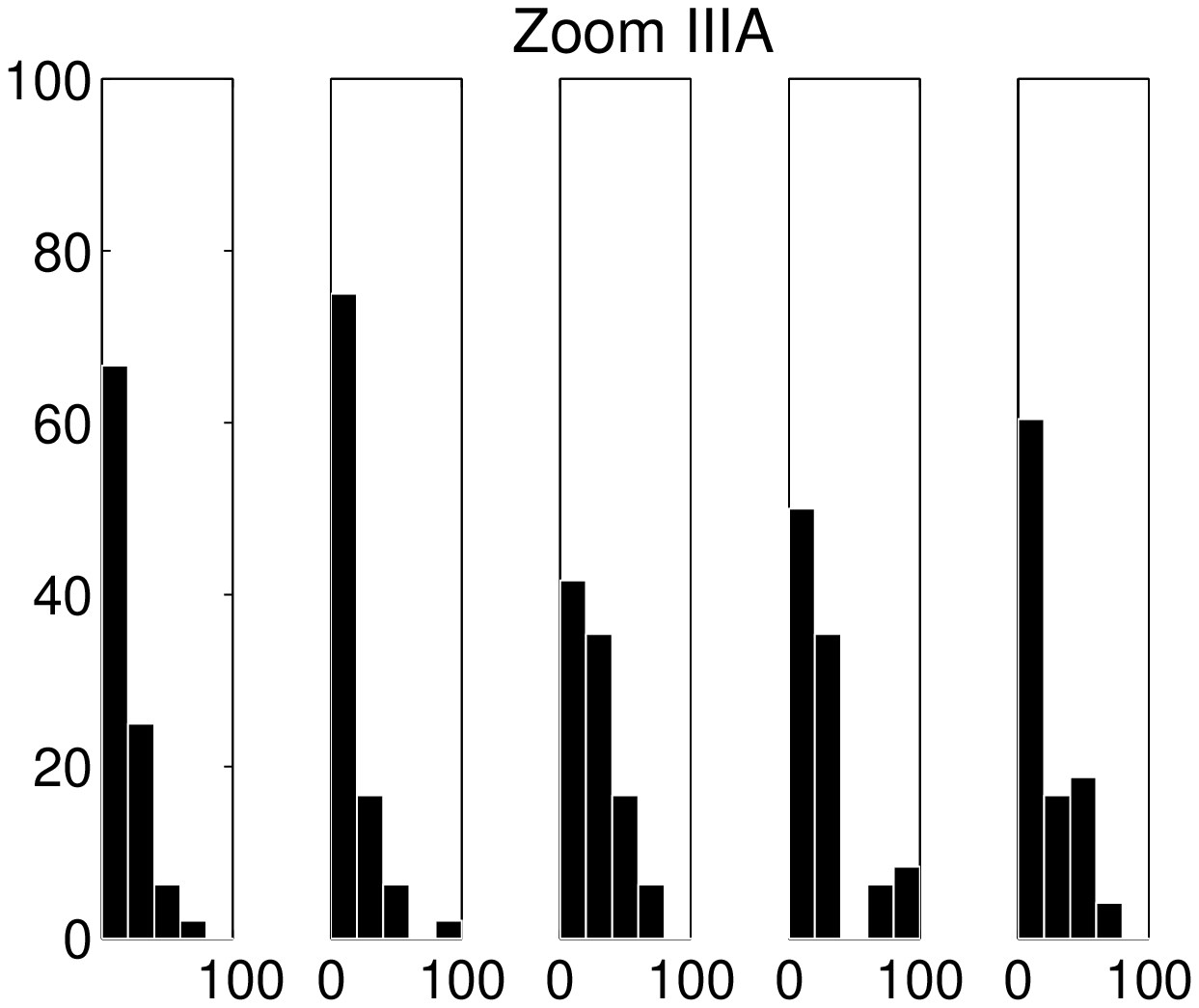}
\caption{
Histograms for Monte Carlo simulations versus each component of $f_0(1370)$ obtained
over subsets $S_\mathrm{IA}$ [relation (\ref{SIA}); the left set of figures], $S_\mathrm{IIA}$ [relation (\ref{SIIA}); the middle  set of figures] and $S_\mathrm{IIIA}$ [relation (\ref{SIIIA}); the right  set of figures].
  Each set consists of five figures,
the vertical axes represent the percentages of Monte Carlo simulations, the horizontal axes (in each set)
from left to right respectively represent the percentages of
		components
		${\bar u}{\bar d} u d$, $({\bar d}{\bar s} d s + {\bar s}{\bar u} s u)
		/\sqrt{2}$, $s {\bar s}$,
		$(u {\bar u}  + d {\bar d} )/\sqrt{2}$, and
		glue.
		Each component is divided into five percentage intervals (0-20\%, 20\%-40\%, etc).
        For example, the first figure on the left shows that more than 90\% of Monte Carlo simulations in zoom IA  resulted in estimating the ${\bar u}{\bar d} u d$ component of $f_0(1370)$ to be below 20\%, and less than 10\% of the simulations resulted in estimating the  $({\bar d}{\bar s} d s + {\bar s}{\bar u} s u)
		/\sqrt{2}$ component between 20\%$-$40\%, etc.
}
\label{F_hist_zoom}
\end{center}
\end{figure}

\begin{figure}[h]
\begin{center}
\vskip .75cm
\epsfxsize = 5.5 cm
 \epsfbox{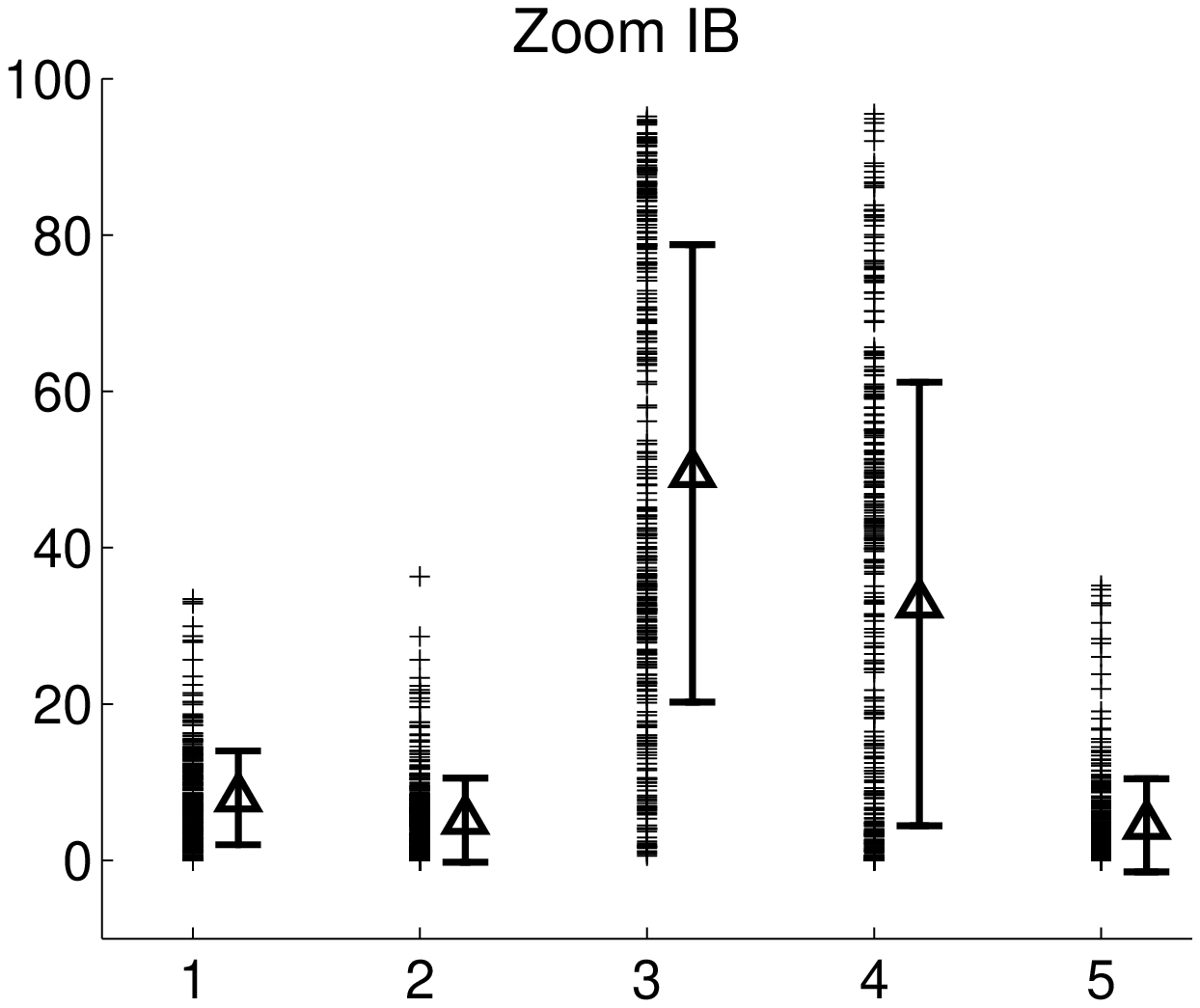}
 \hskip 1 cm
\epsfxsize = 5.5 cm
 \epsfbox{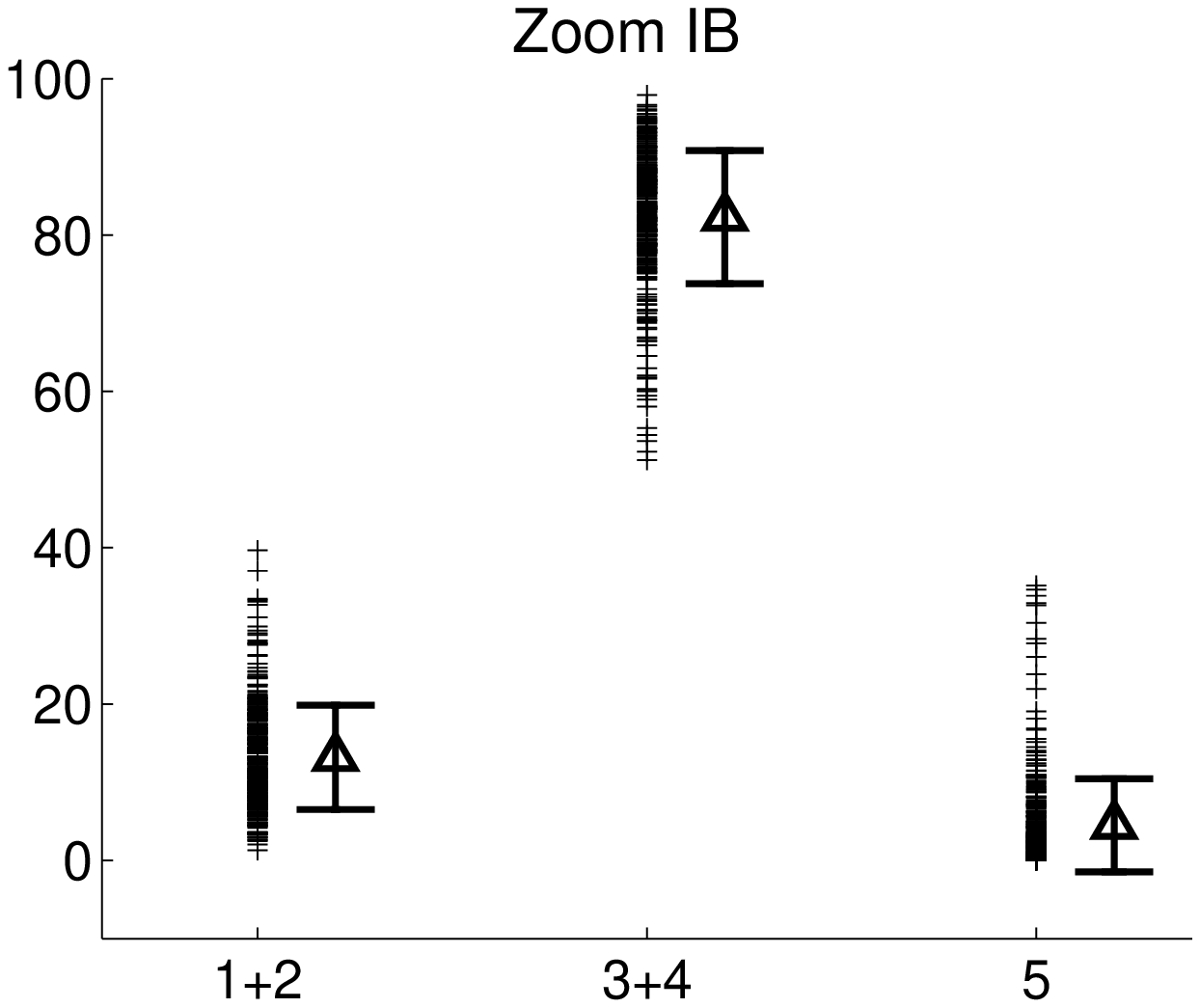}
\caption{Percentages of the quark and glue components of $f_0(1370)$ obtained over subsets $S_\mathrm{IB}$ [relation (\ref{SIB})].
  Components 1 to 5 respectively represent
		${\bar u}{\bar d} u d$, $( {\bar d}{\bar s} d s + {\bar s}{\bar u} s u)
		/\sqrt{2}$, $s {\bar s}$,
		$(u {\bar u}  + d {\bar d})/\sqrt{2}$, and
		glue.  The percentages of total four-quark (1+2), total quark-antiquark (3+4) and glue (5) are also given (right figure).
}
\label{F_comps_zoom2}
\end{center}
\end{figure}
\begin{figure}[h]
\begin{center}
\vskip .75cm
\epsfxsize = 8.0 cm
 \epsfbox{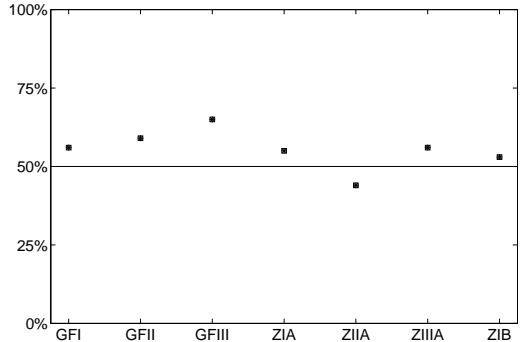}
\caption{Comparing the $s {\bar s}$ and $n {\bar n}$ components of $f_0(1370)$.   Vertical axis shows the percentage of simulations in which the $s {\bar s}$ component is greater than the $n {\bar n}$ component. The horizontal axis gives from left to right the three global fits followed by their zooms A and zoom IB.}
\label{F_ss_vs_nn}
\end{center}
\end{figure}

\section{Summary}

In this work we studied the internal substructure of $f_0(1370)$ within a framework that is designed to explore global properties of  scalar mesons below and above 1 GeV based on various underlying mixings among two- and four-quarks as well as a glue  component.   The framework is based on a nonlinear chiral Lagrangian model with two scalar meson nonets (a four-quark nonet and a quark-antiquark nonet) and a scalar glueball.    This framework has been used in series of prior works (refs. \cite{3flavor}-\cite{Mec}) on the subject and has given a coherent description of various low-energy experimental data.     We investigated the 14d parameter space of the model by performing global fits to the properties of scalars.    We determined sets of points that give an acceptable overall description of experimental data, and thereby computed the quark and glue components of $f_0(1370)$.    It was observed that this state is dominantly a quark-antiquark state (particularly $s{\bar s}$)  with some remnant of four-quark and glue components.   We tested the robustness of the conclusions made, by further zooming in on the properties of $f_0(1370)$ within the global context where the model has an overall agreement with experiment for all scalars below and above 1 GeV.  It was observed that while individual components somewhat vary the overall features remains intact, namely that the $s{\bar s}$ component of $f_0(1370)$ is dominant.

Since the Lagrangian for the $I=0$ scalars (\ref{L_int_I0}) studied in this work is constrained by the Lagrangian of the $I=1/2,1$ states studied in \cite{Mec},  we test the stability of the results when the $I=1/2,1$ Lagrangian parameters are relaxed and included in our global fit.    Overall, we find that there is no noticeable change in the results and that the conclusion for the substructure of $f_0(1370)$ remains the same.    The additional flexibility can be used to further investigate zoom B discussed in previous section.   We find that global set III contains a point that describes all the experimental inputs for $f_0(1370)$ given in Table \ref{quantities2}, i.e.
\begin{equation}
S_\mathrm{IIIB} = \left\{
 p \,| \, p \in S_\mathrm{III} \,:\, m[f_0(1370)], \Gamma^3_{\pi\pi}, \Gamma^3_{KK}\,\, {\rm are \,\, within \,\, their \,\, experimental \,\, ranges } \right\}.
 \label{SIIIB}
\end{equation}
For this point the components of $f_0(1370)$ are shown in Fig. \ref{F_ZoomIIIB} and further confirm the results found in this work for the substructure of this state.

Finally, we point out the measurement of $B_s \rightarrow J/\psi \pi^+ \pi^-$  decay by the LHCb \cite{BstoJpsi_LHCb} where it is reported that  $B_s \rightarrow J/\psi f_0(1370)$ is ``firmly established.''     Belle also has reported the same decay \cite{Belle}. Since this decay proceeds through production of an $s {\bar s}$ pair shown in the schematic diagram of Fig. \ref{Bs}, we interpret this experimental result as some support for our prediction of a significant $s {\bar s}$ component in $f_0(1370)$ in the present study.

\begin{figure}[h]
\begin{center}
\vskip .75cm
\epsfxsize = 5.5 cm
 \epsfbox{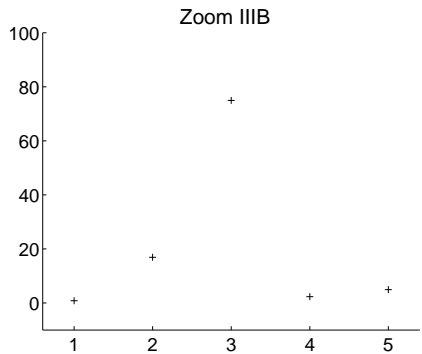}
\caption{Percentages of the quark and glue components of $f_0(1370)$ obtained over subset $S_\mathrm{IIIB}$ [relation (\ref{SIIIB})].
  Components 1 to 5 respectively represent
		${\bar u}{\bar d} u d$, $( {\bar d}{\bar s} d s + {\bar s}{\bar u} s u)
		/\sqrt{2}$, $s {\bar s}$,
		$(u {\bar u}  + d {\bar d})/\sqrt{2}$, and
		glue.}
\label{F_ZoomIIIB}
\end{center}
\end{figure}
\begin{figure}[h]
\begin{center}
\vskip .75cm
\epsfxsize = 7cm
 \epsfbox{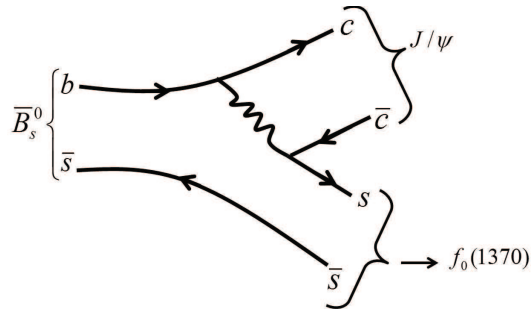}
\caption{Production of $f_0(1370)$ in decay $B_s \rightarrow J/\psi \pi^+ \pi^-$ reported by LHCb \cite{BstoJpsi_LHCb}.   The decay proceeds via production of an $s {\bar s}$ pair and is consistent with the prediction of a large $s {\bar s}$ component in the substructure of $f_0(1370)$ presented in this work.}
\label{Bs}
\end{center}
\end{figure}

\section*{Acknowledgments}
\vskip -.5cm
We are happy to thank S.M. Zebarjad for helpful discussions. A.H.F. wishes to thank J. Schechter for many helpful related discussions, as well as the Physics Dept. of Shiraz University for its hospitality in Summer of 2012 where this work was initiated.

\appendix

\section{Brief review of the mixing mechanism for  $I=1/2$ and $I=1$ scalar states}
Although the scalar mesons above 1 GeV are expected to be close to the conventional quark-antiquark mesons \cite{pdg},  a close look at their properties reveals that this expectation may not be completely supported by experiment.   For example, if $K_0^*(1430)$ and $a_0(1450)$ belong to the same quark-antiquark scalar meson nonet above 1 GeV, then it is rather surprising that  $K_0^*(1430)$, an isodoublet with one strange quark is lighter than the isotriplet member $a_0(1450)$ that should not have any strange quarks in a pure quark-antiquark nonet.   The masses of these two states are reported in   in PDG \cite{pdg}:
 \begin{equation}
m \left[ a_0(1450) \right] = 1474 \pm 19 \hskip .2cm {\rm MeV} > m
\left[ K_0^*(1430) \right] =
1425 \pm 50  \hskip .2cm {\rm MeV}
\label{a0k0_mass_exp}
\end{equation}
which shows that the central value of the mass of isotriplet is about 50 MeV higher than the mass of isodublet.   Even if  we take the  experimental uncertainties into account, which allows the masses to be comparable or get in the right order, still it does not make up for the quark model expectation in which the isodoublet is noticeably heavier than the isotriplet (for example, for the cases of the tensor and axial vector nonets, also p-wave nonets,  the isodoublet is about 100 MeV heavier than the isotriplet).      Also some of the decay ratios of $a_0(1450)$ and $K_0^*(1430)$ do not quite agree with the pattern that  one would expect if these two states were members of a   pure quark-antiquark nonet.
These decay ratios from PDG \cite{pdg} can be compared with the SU(3) predictions (given in parenthesis):
 $
 {\Gamma \left[ a_0^\mathrm{total} \right]}/
{\Gamma\left[K_0^*\rightarrow\pi K\right] } =
0.98 \pm 0.34
\hskip .2cm (1.51)$,
$
\Gamma \left[ a_0\rightarrow K{\bar K} \right]
/
\Gamma\left[a_0\rightarrow\pi \eta\right]  = 0.88 \pm 0.23
\hskip .2cm  (0.55)
$
and
$
\Gamma \left[ a_0\rightarrow \pi\eta' \right]/
\Gamma\left[a_0\rightarrow\pi \eta\right]
$
$
= 0.35\pm0.16 \hskip .2cm (0.16).
$
There are other similar deviations discussed in \cite{Mec}.

A natural question would be whether the deviations of experimental data for $a_0(1450)$ and $K_0^*(1430)$ from what is  predicted if these two states were members of a pure quark-antiquark nonet can be understood based on a mixing of this nonet with the four-quark nonet below 1 GeV.     This question was raised in \cite{Mec} and a mixing mechanism was put forward.   This mechanism is based on a simple picture that provides a natural and effective description of the properties of $I=1/2,1$ scalar mesons below and above 1 GeV ($\kappa(900)$,
$K_0^*(1430)$, $a_0(980)$ and $a_0(1450)$) within in a
nonlinear chiral Lagrangian framework (its extensions to the linear sigma model frameworks have been also studied in
\cite{global}-\cite{05_FJS2}).    The mechanism assumes that there are two scalar meson nonets around about 1 GeV (a lighter pure four-quark nonet $N$ and a heavier  pure  quark-antiquark nonet $N'$) and shows that allowing these two nonets to slightly mix with each other leads to a natural description of the properties of $a_0(1450)$ and $K_0^*(1430)$.
The underlying reason that makes this mechanism successful is due to the  reversed mass ordering of the ``bare'' (unmixed) states in the four-quark  nonet $N$ compared to those in the two-quark nonet $N'$ shown in Fig. \ref{mechanism}.  From the lightest to heaviest these four ``bare'' masses are as follows:  The lightest ``bare'' mass is the $I=1/2$ in nonet $N$ which has one strange quark and is therefore lighter than the $I=1$ of this four-quark nonet followed by  the $I=1$
state of nonet $N'$ with no  strange quarks and the heaviest $I=1/2$ ``bare'' state in nonet $N'$ which has one strange quark. Therefore,  in the ``bare'' mass spectrum the two $I=1/2$ states are farthest apart and the two $I=1$ states are closest to each other.    This reverse ordering  turns out to be the magic behind this mechanism.     To see how this works, we first remember the following simple property in small mixing $\epsilon$ of two states of mass $m_2 > m_1$ that results in physical masses ${\tilde m}_2 > {\tilde m}_1$ described by mass matrices
\begin{equation}
\left[
\begin{array}{cc}
m_1^2 & \epsilon\\
\epsilon & m_2^2
\end{array}
\right]
\hskip 1cm
\rightarrow
\hskip 1cm
\left[
\begin{array}{cc}
{\tilde m}_1^2 & 0\\
0 & {\tilde m}_2^2
\end{array}
\right]
\end{equation}
We can easily show
\begin{eqnarray}
{\tilde m}_1^2 &=& m_1^2 - {\epsilon^2 \over \delta} \nonumber \\
{\tilde m}_2^2 &=& m_2^2 + {\epsilon^2 \over \delta}
\end{eqnarray}
where $\delta = m_2^2 - m_1^2$.   When the two ``bare'' masses are degenerate ($m_1$=$m_2$) the physical masses are ${\tilde m}_1^2 = m_1^2 - \epsilon$,  ${\tilde m}_2^2 = m_2^2 + \epsilon$. This shows that when the two ``bare'' states of mass $m_1$ and $m_2$ mix,  the physical masses split away from the ``bare'' masses and that the magnitude of this splitting is inversely proportional to the difference of the ``bare'' masses squared.    Using this property we can see in Fig. \ref{mech_mass}(a) that when the two $I=1$ states mix, since they are closer to each other,  they split more than the two $I=1/2$ states (since $\delta_1 < \delta_{1/2}$), and as a result, a level crossing takes place where the $a_0(1450)$ get pushed above the $K_0^*(1430)$ and therefore we can understand the experimental data based on the  mixing of a  pure quark-antiquark nonet with a four-quark nonet. The case of $\delta_1=0$ is shown in Fig. \ref{mech_mass}(b).
 Also shown in Fig. \ref{mech_mix} is an schematic quark line diagram for the mixing of two states with the same quantum numbers (one in the four-quark nonet $N$ and another one in the two-quark nonet $N'$)  in which a possible rearrangement of quark lines to generate the two mixed physical states can be seen.    In addition to  describing the mass spectrum, the mixing also makes it possible to describe the decay ratios mentioned above.

The Lagrangian density for the $I=1/2,1$ scalars is developed in \cite{Mec}
\begin{eqnarray}
{\cal L}_{\rm mass}^{I=1/2,1} &=&
- a {\rm Tr}(NN) - b {\rm Tr}(NN{\cal M})
- a' {\rm Tr}(N'N') - b' {\rm Tr}(N'N'{\cal M}) -\gamma {\rm Tr} \left( N N' \right),
\label{L_mass_I1}\\
{\cal L}_{\rm int}^{I=1/2,1} &=&
+ A \epsilon^{abc}\epsilon_{def}
N_a^d\partial_\mu\phi^e_b\partial_\mu\phi^f_c
+ C {\rm Tr} (N \partial_\mu \phi) {\rm Tr} (\partial_\mu\phi)
+ A' \epsilon^{abc}\epsilon_{def}
{N'}_a^d\partial_\mu\phi^e_b\partial_\mu\phi^f_c
\nonumber \\
&&+ C' {\rm Tr} (N' \partial_\mu \phi) {\rm Tr} (\partial_\mu\phi),
\label{L_int_I1}
\end{eqnarray}
where ${\cal M} = {\rm diag} (1,1,x)$ with $x$ being the ratio of
the strange to non-strange quark masses, and $a,b,a'$, $b'$, $\gamma$, $A$, $C$, $A'$ and $C'$ are a priori
unknown parameters fixed by experimental inputs on mass and decay properties of $I=1/2$ and $I=1$ states below and above 1 GeV.

\begin{figure}[h]
\begin{center}
\vskip .75cm
\epsfxsize = 8 cm
 \epsfbox{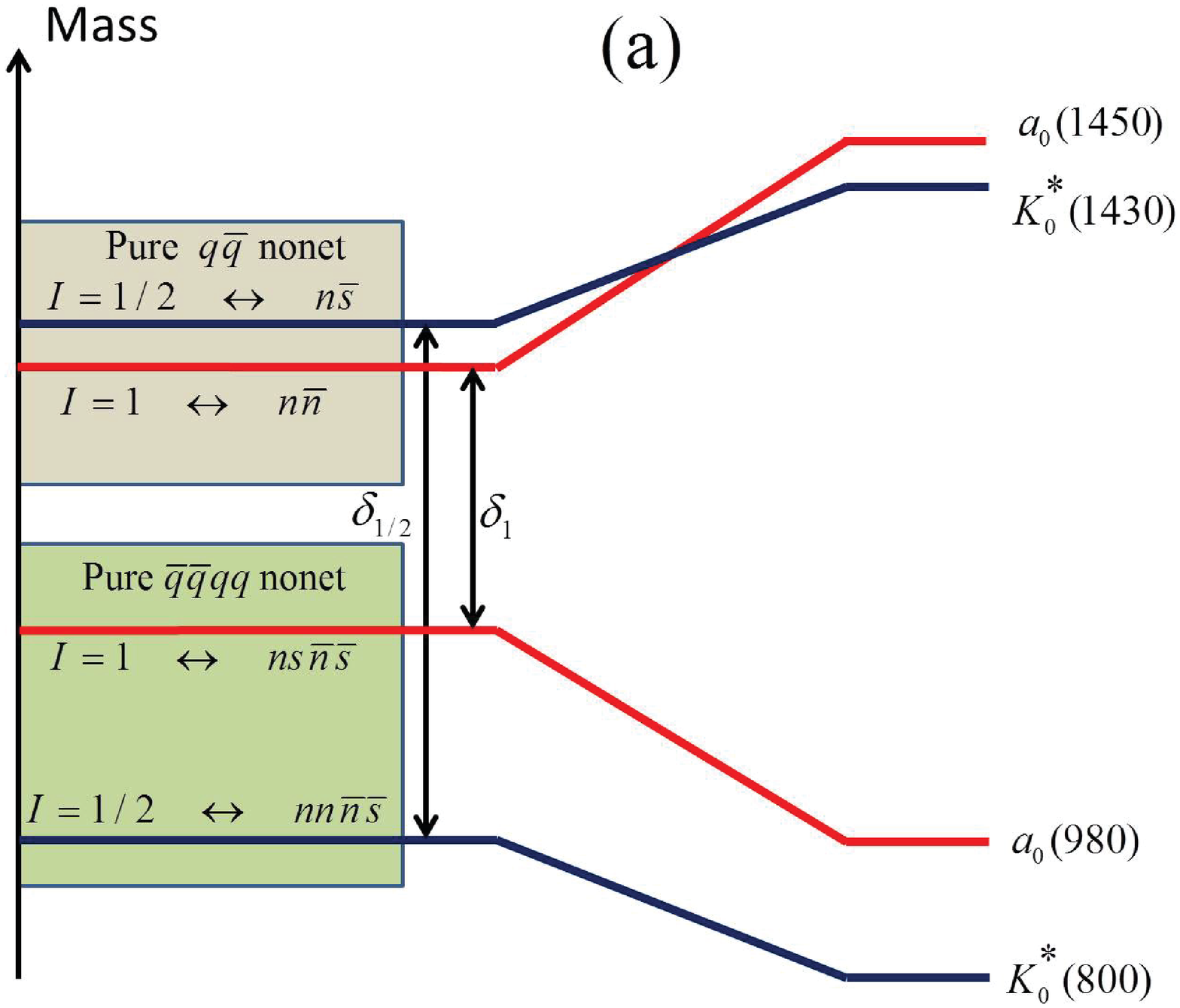}
 \epsfxsize = 8 cm
 \epsfbox{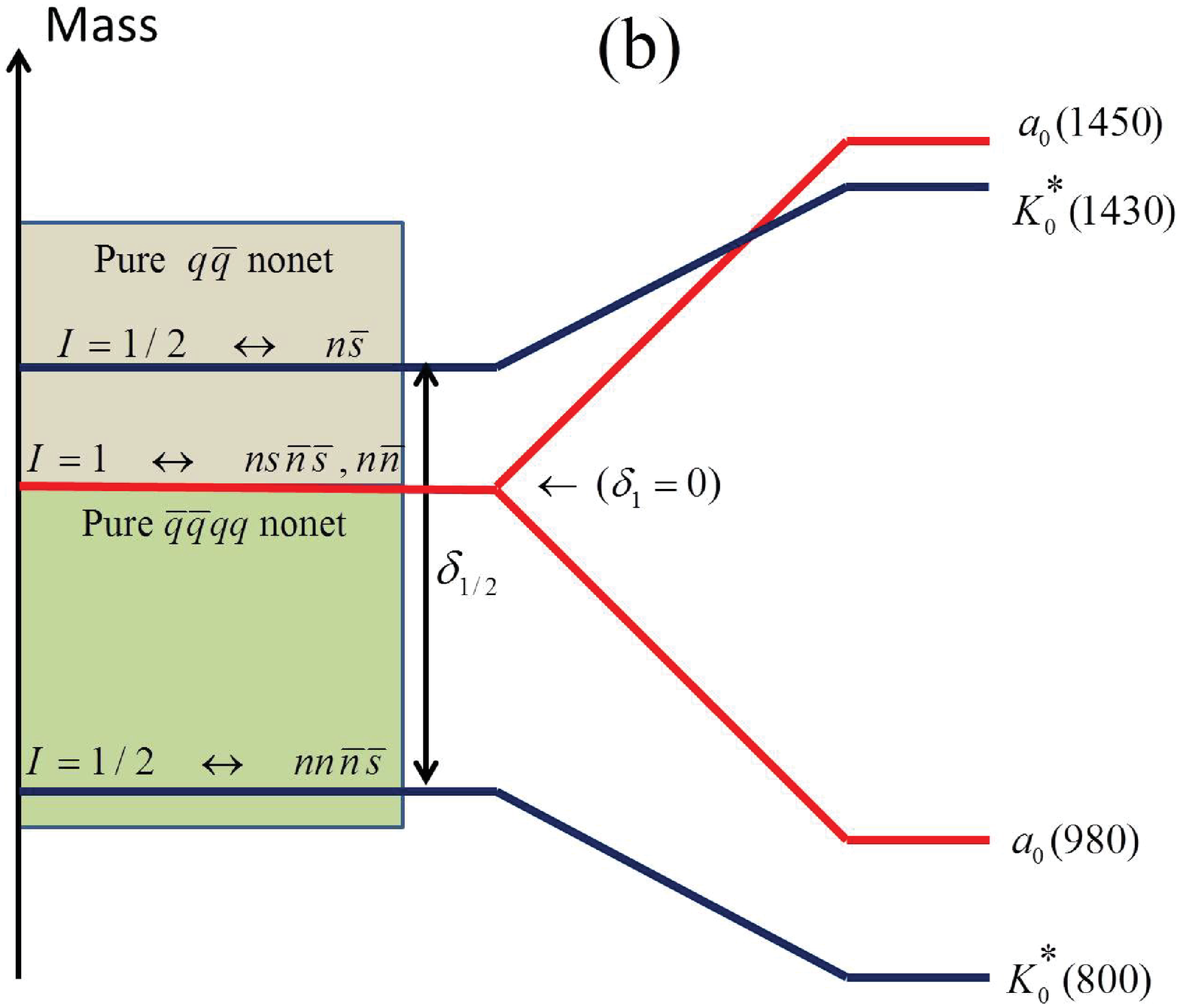}
\caption{Mixing mechanism for isodoublet and isotriplet scalar mesons below and above 1 GeV.     Existence  of a pure four-quark scalar meson nonet beneath a quark-antiquark nonet [Fig. (a)] results in a mass spectrum for $I=1/2,1$ states where the two isotriplets are closer to each other than the two isodoublets (i.e. $\delta_1 < \delta{1/2}$). This is due to the inverted mass spectrum in a four-quark nonet compared to that of a quark-antiquark nonet.  Allowing the states with the same quantum numbers to mix leads to a level repulsion in which the isotriplets split more than the isodoublets and consequently a level crossing occurs that results in $a_0(1450)$ becoming heavier than the $K_0^*(1430)$.  This mechanism also works when the two isotriplets are degenerate in mass [Fig. (b)].   The  mechanism developed in ref. \cite{Mec}, in the leading order,  favored situation (b).   The level repulsion also shows that the light scalar mesons below 1 GeV get pushed down and become lighter than expected.
 \label{mechanism}    }
\label{mech_mass}
\end{center}
\end{figure}
\begin{figure}[h]
\begin{center}
\vskip .75cm
\epsfxsize = 10 cm
 \epsfbox{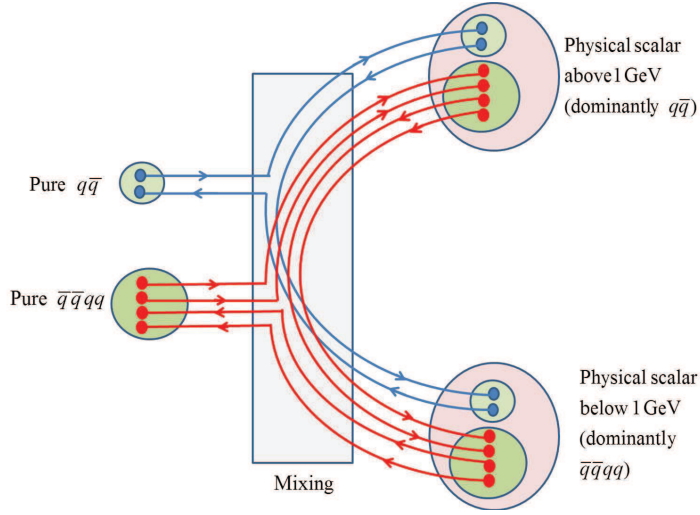}
\caption{
Schematic diagram  for a possible rearrangement of the quark lines in the mixing process.}
\label{mech_mix}
\end{center}
\end{figure}

Note that the last term in (\ref{L_mass_I1}) which induces the mixing between the two- and the four-quark nonets $N'$ and $N$ can be rewritten as
\begin{eqnarray}
-\gamma {\rm Tr} \left( N N' \right) &=& - \gamma N_a^d {N'}_d^a \nonumber\\
& \propto& \left( Q_a {\bar Q}^d\right) \left( q_d {\bar q}^a \right)
=\left( \epsilon_{abc}{\bar q}^b{\bar q}^c \right)
\left( \epsilon^{def}q_e q_f \right)
\left( q_d {\bar q}^a \right) \nonumber\\
&=&\epsilon^{def} \epsilon_{abc}   \left( q_d {\bar q}^a \right) \left( q_e {\bar q}^b \right) \left( q_f {\bar q}^c \right)
={1\over 6} \det \left[ N'(q) \right].
\end{eqnarray}
where ${N'}_a^b (q) = q_a {\bar q}^b$.
The determinant structure is similar to the contribution of instantons to the scalar channel discussed in \cite{Klempt_1995,Minkowski_99,Dmitrasinovic_96}.

With
\begin{equation}
\begin{array}{cclcccl}
m^2_{a_0} &=& 2 (a + b), & &  m^2_{a'_0} & = & 2 (a' + b'),\\
m^2_{K_0} &=& 2a + (1+x) b, & &  m^2_{K'_0} & = & 2 a' + (1+x) b',
\end{array}
\end{equation}
it is shown in \cite{Mec} that
\begin{equation}
m_{a_0}=m_{a'_0}=1.24 \hskip .1cm{\rm GeV},\hskip .2cm
m_{K_0}=1.06 \hskip .1cm{\rm GeV}, \hskip .2cm
m_{K'_0}=1.31 \hskip .1cm{\rm GeV}, \hskip .2cm
\gamma = 0.58 \hskip .1cm {\rm GeV^2}.
\label{a0k0_m_bare}
\label{bare_masses}
\end{equation}

The interaction coupling constants are also found \cite{Mec} from various decay widths of isodoublets and isotriplet states
\begin{equation}
A=1.19\pm 0.16 \hskip .1cm {\rm GeV}^{-1}, \hskip 0.2cm
A'=-3.37\pm 0.16 \hskip .1cm  {\rm GeV}^{-1}, \hskip 0.2cm
C=1.05\pm 0.49 \hskip .1cm  {\rm GeV}^{-1}, \hskip 0.2cm
C'=-6.87\pm 0.50 \hskip .1cm  {\rm GeV}^{-1}. \hskip 0.2cm
\label{AC_values}	
\end{equation}

Further details can be found in \cite{Mec}.

\end{document}